\documentclass{article}

\usepackage[english]{babel}
\usepackage[letterpaper,top=2cm,bottom=2cm,left=3cm,right=3cm,marginparwidth=1.75cm]{geometry}
\usepackage{amsmath}
\usepackage{amsfonts} 
\usepackage{graphicx}

\usepackage[colorlinks=true, allcolors=blue]{hyperref}

\usepackage[style=vancouver, url=false]{biblatex}
\AtEveryBibitem{
  \clearfield{month}
}
\renewbibmacro*{volume+number+eid}{%
  \printfield{volume}%
  \setunit*{\addnbspace}
  \printfield{number}%
  \setunit{\addcomma\space}%
  \printfield{eid}}
\DeclareFieldFormat[article]{number}{\mkbibparens{#1}}

\title{Emergence of metastability in frustrated oscillatory networks: the key role of hierarchical modularity}
\author{Enrico Caprioglio and Luc Berthouze \\ \footnotesize{Department of Informatics, University of Sussex, BN1 9RH Brighton, UK} \\ \footnotesize{\{E.Caprioglio, L.Berthouze\}@sussex.ac.uk}}

\begin{document}
\maketitle

\begin{abstract}
Oscillatory complex networks in the metastable regime have been used to study the emergence of integrated and segregated activity in the brain, which are hypothesised to be fundamental for cognition. Yet, the parameters and the underlying mechanisms necessary to achieve the metastable regime are hard to identify, often relying on maximising the correlation with empirical functional connectivity dynamics. Here, we propose and show that the brain's hierarchically modular mesoscale structure alone can give rise to robust metastable dynamics and (metastable) chimera states in the presence of phase frustration. We construct unweighted $3$-layer hierarchical networks of identical Kuramoto-Sakaguchi oscillators, parameterized by the average degree of the network and a structural parameter determining the ratio of connections between and within blocks in the upper two layers. Together, these parameters affect the characteristic timescales of the system. Away from the critical synchronization point, we detect the emergence of metastable states in the lowest hierarchical layer coexisting with chimera and metastable states in the upper layers. Using the Laplacian renormalization group flow approach, we uncover two distinct pathways towards achieving the metastable regimes detected in these distinct layers. In the upper layers, we show how the symmetry-breaking states depend on the slow eigenmodes of the system. In the lowest layer instead, metastable dynamics can be achieved as the separation of timescales between layers reaches a critical threshold. Our results show an explicit relationship between metastability, chimera states, and the eigenmodes of the system, bridging the gap between harmonic based studies of empirical data and oscillatory models.
\end{abstract}

\section{Introduction}\label{sec:introduction}
The macroscopic dynamics of the human brain are characterized by a balance and flexible switching between integrated and segregated activity \cite{Fox2007}. The coexistence of both segregative and integrative tendencies is thought to be essential for healthy brain functioning and cognitive processing \cite{Tognoli2014, Capouskova2023, LpezGonzlez2021, Wang2021}. The flexible switching between those tendencies among the systems' parts is the defining feature of complex dynamical systems in the metastable regime \cite{Kelso2013}. In the last two decades, various indexes of metastability have been proposed to quantify these dynamical behaviours from empirical data \cite{Hancock2023}. Recently, measures of metastability have been successful in characterizing changes in the brain's metastable-like dynamical features in the presence of pathological or pharmacological alterations \cite{Lee2018, Lord2019, Hancock2023schizophrenia}, brain injury \cite{Hellyer2015}, the aging brain \cite{Cabral2017Aging} (as well as the aging rat brain \cite{deAlteriis2024}), and after brain stimulation \cite{Bapat2024}.
\newline

Motivated by these empirical observations, numerous whole-brain dynamical models have been employed to reproduce the metastable features of macroscopic brain dynamics \cite{Cabral2014review, Hansen2015}. A whole-brain model consists of a complex dynamical system in which each of the constituent elements interacts with other elements according to the structural connectivity of real brains. The most frequently used indexes of metastability are the variation of the Kuramoto Order Parameter (KOP) \cite{Shanahan2010} and the variation of the leading eigenvector of the instantaneous phase locking matrix and its eigenvalue \cite{Cabral2017Aging, deAlteriis2024} (see review \cite{Hancock2023} for an in-depth discussion). Whole-brain oscillatory based models, in particular, allow for an extensive exploration of the structural and dynamical parameter space, enabling researchers to identify key parameters that dictate the models' behaviours in the metastable regime \cite{Cabral2011, Deco2017, Torres2024}. A large number of key control parameters have been suggested to play a vital role in the emergence of metastable-like features. For instance, heterogeneity in couplings, delays, and connectivity has been shown to be crucial for the emergence of metastable states that synchronize at frequencies lower than the average oscillator frequency \cite{Cabral2022}. Turning to the role of structure, graph theoretical properties of connectomes have been shown to alter indexes of metastability \cite{Va2015} and similar measures of functional complexity \cite{ZamoraLpez2016}.
\newline

In these models, indexes of metastability have been hypothesised to peak when the correlation with empirical dynamical functional connectivity (dFC) is maximised \cite{Deco2016} (however, see also \cite{Pope2023}). Thus, most studies primarily focus on the properties of a system in the metastable regime and how the system's parameters change these observations. However, the criteria that should be satisfied by an oscillatory system to achieve the metastable regime, and the mechanisms underpinning metastability, are not fully understood. As suggested in \cite{Cabral2022}, understanding the mechanisms that give rise to the emergence of metastable states is essential to be able to make predictions about the appearance and duration of specific metastable modes, and eventually to design possible therapeutic interventions.
\newline

In dynamical systems theory, metastable-like dynamics have been observed in various systems of oscillators. Here, metastability is usually associated with instability of chimera states: symmetry breaking states characterised by the coexistence of coherent and incoherent patterns in systems of phase-lagged identical oscillators \cite{Abrams2004, Panaggio2015, Haugland2021}. For instance, in \cite{Shanahan2010}, metastable chimera states arise as a result of winnerless competition between modules of identical oscillators to join the coalition of synchronized modules. Similar metastable behaviours can be found in \cite{Bick2018} due to the addition of higher-order interactions. Chaos, turbulence, and other dynamics with metastable-like features have also been shown to arise in the case of heterogeneous couplings and phase lags in the two-population model \cite{Bick2018chaos}. Thus, these studies suggest that the interplay between modular structures, coupling heterogeneities, and phase frustration is necessary for metastable chimera states to arise. The effects of non-local hierarchical topologies on chimera states have also been investigated. In these settings, the structural symmetries and the clustering coefficient have been found to be important parameters for the emergence of chimera states in networks of Van der Pol oscillators \cite{Ulonska2016}. Closer to the approach taken in this study, authors in \cite{Makarov2019} analyzed a system of Kuramoto-Sakaguchi oscillators composed of subnetworks connected via hub nodes placed on a ring. The authors showed how the size and connectivity of the subnetworks promote a competition mechanism between scales which results in an expansion of the range of parameters within which chimera states arise when compared to the case without subnetworks. Hence, hierarchical network properties seem to promote the emergence of chimera states, but their role in achieving the metastable regime and the specific mechanisms enabled by hierarchical structures remain unclear.
\newline

Some insights have come from computational neuroscience studies that explored in more detail effects of the hierarchically modular structure of the brain \cite{Meunier2010}, such as Griffith phases \cite{Rubinov2011, Moretti2013, dor2015}. In particular, recent research suggested that including the brain's mesoscale connectivity in whole-brain oscillatory models can help us identify and understand the mechanisms underlying metastable dynamics. In \cite{Mackay2023}, the authors generated spatially constrained random networks to approximate the mesoscale neocortex, allowing the implementation of heterogeneous phase delays in a model of non-identical Kuramoto oscillators. The authors observed how the topology of these constrained networks widens the range of critical couplings within which metastable behaviours arise, analogous to Griffith phases. In \cite{Villegas2014} instead, the authors studied in detail real and synthetic hierarchically modular brain networks of oscillators in the absence of phase frustration. Whilst it is well known that the hierarchical network structure affects the dynamics of oscillatory systems in the path towards global synchronization \cite{Arenas2006, Arenas2007, Villegas2022}, in \cite{Villegas2014} the authors observed that modules of identical units at the bottom of the hierarchy synchronize first (fast timescale dynamics). However, synchronization can be lost as interactions with other modules become more significant at longer timescales, giving rise to transient metastable states on the path towards synchronization. Thus, these observations highlight how the slow dynamics associated with the lowest eigenvalues of the graph Laplacian affect the faster timescale dynamics that dictate the oscillators' behaviour within lower-layer modules. This study, however, did not consider the effect of phase frustrations, such as phase lags or phase delays, and whether robust metastable and chimera states can persist, preventing global synchronization even in the case of identical oscillators.
\newline

Taken together, these studies suggest that modular or hierarchically modular structures, combined with either higher-order (non-additive) interactions, or the heterogeneity of both couplings and delays, are crucial to achieve the metastable regime (but also see the case for nonlinear couplings \cite{Haugland2015, Haugland2021}). In this work, we investigate the more parsimonious hypothesis that the brain’s hierarchically modular mesoscale structure alone can give rise to robust chimera states and modulate metastable dynamics in the presence of phase-frustration. Under the assumption that dFC patterns in real brains are supported by the brain's mesoscale structural connectivity, our hypothesis is motivated by the fact that dFC patterns can be constructed using the spectral information of structural brain networks \cite{Atasoy2016, Atasoy2017}. Inspired by the network structures used in \cite{Villegas2014, ZamoraLpez2016, Kang2019, Fusc2023}, we construct 3-layer hierarchical networks using a variation of the nested Stochastic Block Model (nSBM) \cite{Peixoto2014}. Our variation of the nSBM is parametrized by two quantities: the average degree of the network, and a structural parameter controlling the ratio of connections between and within blocks at high hierarchical layers while keeping the average degree of the network conserved and homogeneous. In the absence of coupling and oscillator frequency heterogeneities, we show how the Kuramoto-Sakaguchi dynamics on these networks display robust metastable and chimera states at different hierarchical layers depending on the mesoscale structure of the network. Through explaining the mechanisms underpinning the emergence of metastable and chimera states we elucidate an explicit relationship between these states and the eigenmodes of the graph Laplacian. In particular, we show how the system we investigated may be reduced to the classic two-population model for a specific range of the structural parameter via the Laplacian Renormalization Group (LRG) flow. Our results suggest that, while the slow eigenmodes determine the functional organization in higher coarse-grained layers, the spectral gap between fast and slow eigenmodes affects the stability of cluster synchronization in the lower fine-grained layers. We conclude by pointing towards possible future extensions of this work to further bridge the gap between harmonic based studies of empirical dFC and oscillatory based whole-brain models.

\section{Methods}\label{sec:methods}
\subsection{Hierarchical Network Model Choice}\label{sec:ModelChoice}
To test our hypothesis, we seek to construct hierarchically modular networks while avoiding the presence of structural heterogeneities such as network motifs, hubs and the rich club, as well as non-homogeneous degree distributions, all of which have already been studied in \cite{Ulonska2016, Krishnagopal2017, ZamoraLpez2016}. Hence, we seek to construct hierarchical networks with a single control parameter that smoothly varies the mesoscale organization of the network while maintaining the average degree of the network fixed and homogeneous across all vertices. Critical mesoscale structures, which, in the context of information-diffusion and synchronization dynamics, are identified as structures that emerge at a characteristic timescale in the dynamical process, are an important feature of hierarchically modular networks such as the brain \cite{Villegas2022}. Such structures naturally arise in community-structured networks such as the Stochastic Block Model (SBM) and its nested variations (nSBM) used in various computational neuroscience studies such as \cite{Villegas2014, Kang2019, Villegas2022, Fusc2023}. Therefore, nSBMs are a good starting point for our study. In the next subsection, we introduce a single-parameter nSBM variation, allowing us to smoothly vary the system from a SBM to a nSBM for a given desired average degree.

\subsection{Variation of the nSBM}\label{sec:variationSBM}
\begin{figure}[h]
    \centering
    \includegraphics[width=0.95\textwidth]{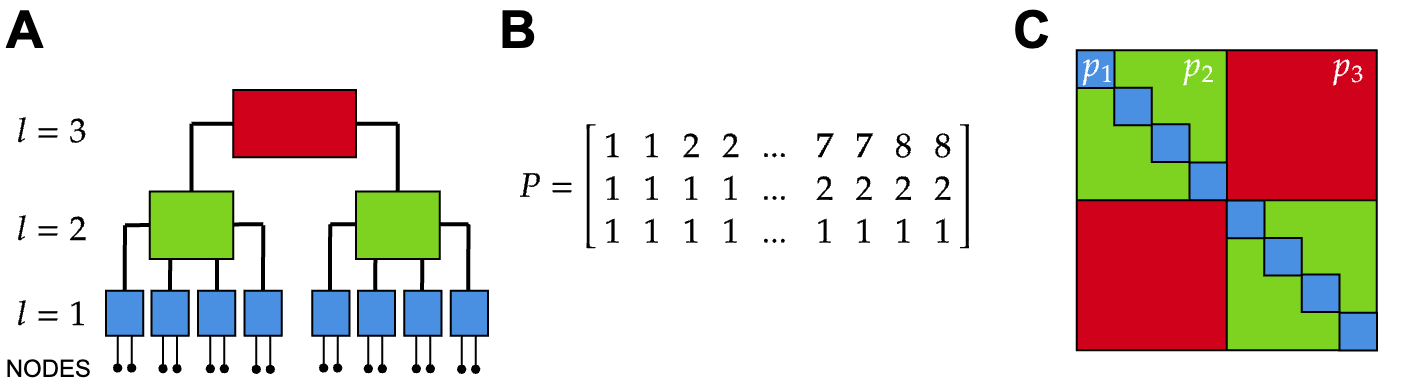}
    \caption{Example of a $3$-layer hierarchical network. (A) At each layer $l$, we define a SBM with $B_l$ blocks containing $n_{l}$ blocks from the layer below (or nodes if $l=1$). The last layer is fixed to be a single block, $B_3 = 1$, containing $n_3 = 2$ blocks from layer $2$. In this example in layer $1$ (blue blocks) we have $B_1=8, n_1=2$, in layer $2$ (green blocks) we have $B_2=2, n_2=4$, in layer $3$ (red block) we have $B_3=1, n_3=2$. (B) Corresponding partition matrix. (3) Using our constraints, the $3$-layer hierarchical network is fully parametrized by the connection probabilities $p_i,\,i=1,2,3$ (blue, green, red).}
    \label{fig:diagram1}
\end{figure}

\begin{figure}[h]
    \centering
    \includegraphics[width=0.95\textwidth]{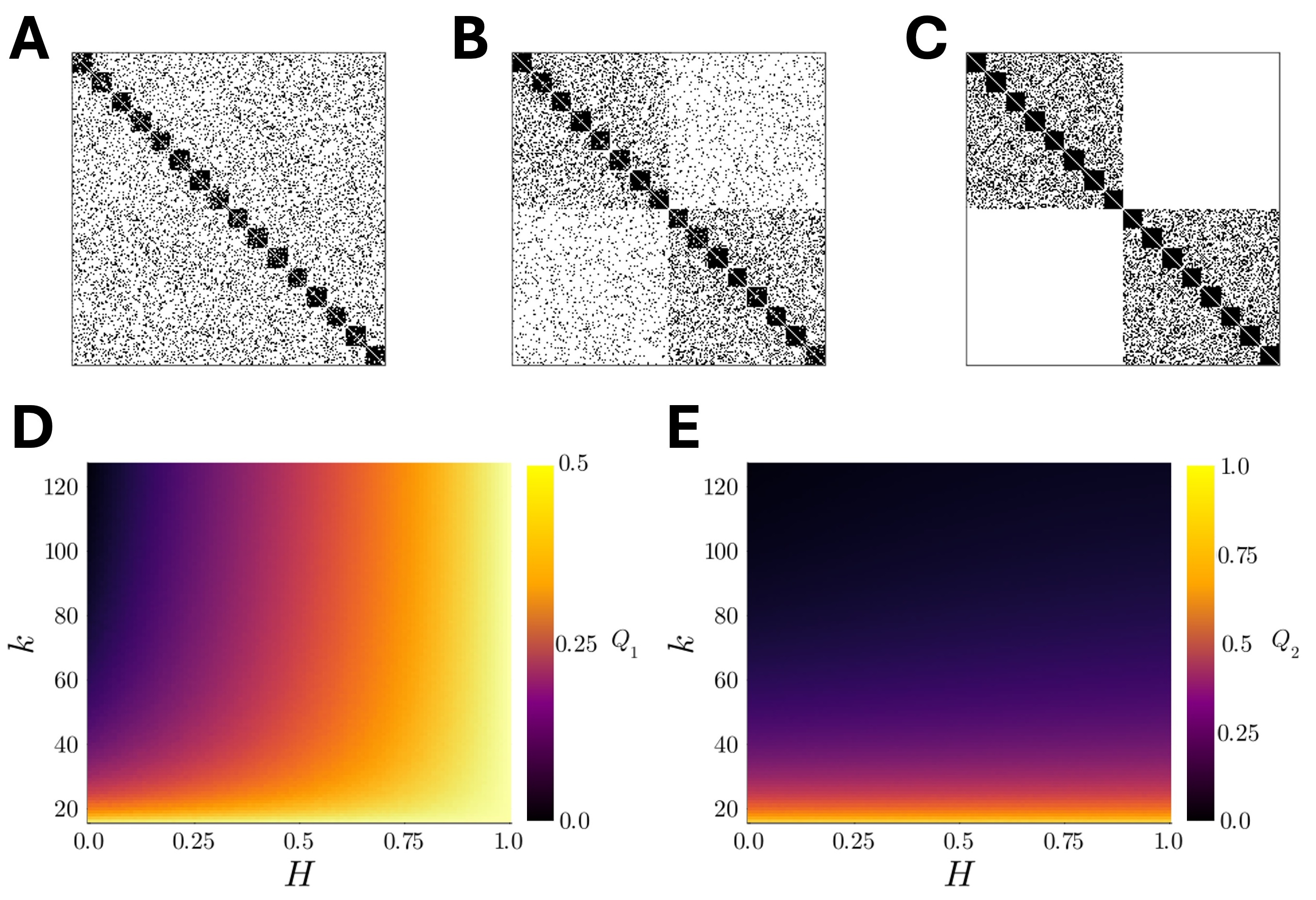}
    \caption{(top) Example adjacency matrices for $H = 0.0$ (\textbf{A}) $H = 0.5$ (\textbf{B}) and $H = 1$ (\textbf{C}). Each network was generated using $n_1 = 16,\,n_2=8$ and exemplifying average degree $k = 51.2$. Note how for $H=0.0$ the generated network is similar to a SBM while for $H=1.0$ the two populations are decoupled. (bottom) Modularity index $Q_l$ calculated over partitions $\{P_{l,i}\,|\,i\in\mathcal{V}\}$ for $l = 1$ (\textbf{D}) and $l=2$ (\textbf{E}) as a function of $H$ and $k$.}
    \label{fig:ExampleMatrices}
\end{figure}

Given an arbitrary partition $P$ of $N=|\mathcal{V}|$ nodes, where $\mathcal{V}$ is the set of nodes, a SBM is defined by assigning a probability of connection between all pairs of nodes in terms of $P$. Starting from an initial node partition into $B_1$ blocks (layer $l=1$), we obtain a $L$-layer nSBM \cite{Peixoto2014} by defining a new SBM in each layer $l'>l$ in terms of the partition at layer $l'-1$ (\hyperref[fig:diagram1]{Fig. 1A}). The last layer $L$ is defined as a single block for convenience. This process results in a partition matrix $P \in \mathbb{N}^{L \times N}$ (\hyperref[fig:diagram1]{Fig. 1B}). Each column of the partition matrix $P$ assigns a node to a specific block in each layer $l$, where each row of $P$ corresponds to the partition in layer $l$. Thus, the system is parametrized by assigning a probability of connection to all $N(N-1)$ possible connections in terms of the partition matrix of dimension $L \times N$.  We may reduce the dimension of the $[L \times N]$-dimensional parameter space by choosing specific probability of connection rules that depend on the nodes' block membership in each layer.
In our variation of the nSBM, see \hyperref[fig:diagram1]{Fig. 1C}, we introduce the following constraints:
\begin{itemize}
    \item each block in layer $l$ contains the same number $n_{l}$ of layer $l-1$ blocks;
    \item the probability of connection between two nodes in a block in layer $l$ is the same for all $B_l$ blocks in the same layer;
    \item  in layer $L$ there is only one block containing $N_{L} = 2$ blocks from layer $L-1$.
\end{itemize}
The first two constraints ensure structural homogeneity at layer level for all layers, while the third constraint was chosen to model the brain's two-hemisphere configuration, similar to previous studies \cite{Villegas2014, Kang2019}. We assign non-zero probabilities only for edges between nodes in the same block of any layer. Then, since layer $L$ is a single block, only $L$ probabilities are required given these constraints. For the case $L = 3$, all possible edges are assigned a probability $p_1,\,p_2,$ or $p_3$ as depicted in \hyperref[fig:diagram1]{Fig. 1C}. However, we wish these probabilities to depend on a single control parameter $H\in[0,1]$ such that $p_i \equiv p_i(H),\, i = 1,2,3$. To do this, we add three more requirements:
\begin{itemize}
    \item for communities in layer $1$ to be almost fully connected, i.e., $p_1$ must be as close to unity as possible;
    \item for layers $2$ and $3$ to be identical when $H = 0$. Then, the nSBM reduces to the classical SBM with only intra- and inter-community connection probabilities, i.e., $p_2(0) = p_3(0)$, with $p_1(H)$ reducing to the intra-community connection probability and $p_2=p_3$ reducing to the inter-community connection probability;
    \item for communities in layer $2$ to be completely decoupled for $H = 1$, such that $p_3(H=1) = 0$.
\end{itemize}
These constraints allow to smoothly vary the system configuration from a $2$-layer community structured network to a $3$-layer hierarchically modular network while keeping blocks in layer $1$ always almost fully connected. Finally, we obtain expressions for $p_1(H),\,p_2(H)$ and $p_3(H)$ by requiring the average degree of the network to remain constant with respect to varying $H$ (see derivation in \hyperref[sec:AppendixA]{Appendix A}):
\begin{eqnarray}
    p_1(H) &=& 1 - \left(\frac{1 - H}{2}\right)\left(\frac{n_1 \gamma}{n_1 - 1}\right), \\
    p_2(H) &=& \left(\frac{1+H}{2}\right)\gamma,\\
    p_3(H) &=& \left(\frac{1-H}{2}\right)\gamma,
\end{eqnarray}
where the factor $\gamma\in[0,1]$ in the expressions for $p_2$ and $p_3$ can be set to control the average degree of the network. Given these constraints, the theoretical average degree of the network can be written as
\begin{equation}
    \langle{k}\rangle = (n_1-1)p_1(H) + n_1(n_2 - 1)p_2(H) + n_2n_1(n_3 - 1)p_3(H),
    \label{eqn:AverageDegree1}
\end{equation}
allowing us to find $\gamma$ given a desired average degree $k$. Intuitively, we may see \hyperref[eqn:AverageDegree1]{Eq. 4} as the sum of the connection probabilities of any single row in \hyperref[fig:diagram1]{Fig. 1B}. Additionally, we may also note how this equation shows the structural homogeneity in each layer; each node (block) has the same average degree as any other node (block) in the same layer. After plugging in the expressions for $p_i$ we obtain the following expression for the average degree of the network (see \hyperref[sec:AppendixA]{Appendix A}):
\begin{equation}
    \langle{k}\rangle = n_1 + \gamma(n_1n_2 - n_1) -1, \quad\quad\gamma\in[0,1].
\end{equation}
Hence our model allows to choose an average degree in the range between $n_1-1$ and $n_1n_2-1$, i.e., the maximum number of connections in a layer 1 module and the maximum number of connections in a layer 2 module. This limitation is a result of the imposed model requirements. Note that the total number of nodes is determined by $n_l$, with $N=n_1n_2n_3$. In what follows, we used $n_3=2$, $B_3=1$, and $B_2=2$, to model the two-hemisphere division of the brain and to allow comparison of our results with the $2$-population model analytical results \cite{Abrams2008}. Finally, following the model's constraints, we also have that $B_1 = 2n_2$. Therefore, given the size of blocks in layers $1$ and $2$, specified by $n_1$ and $n_2$, our variation of the nSBM is fully parametrized by $k$ and $H$, which control the average degree and the density of connections between and within blocks in higher hierarchical layers of the resulting network, respectively. In \hyperref[fig:ExampleMatrices]{Fig. 2}, we show three examples of adjacency matrices constructed using our variation of the nested SBM.
The parameter $H$ changes the density of connections within and between the two populations of nodes, while preserving the average degree and the node's degree distribution of the resulting networks. For easier understanding, in the remainder of the manuscript, we will refer to blocks of nodes in layer $2$ as populations, and blocks of nodes in layer $1$ as modules. Then, the subset of nodes belonging to a population $\rho_i,\,i=1,2$ are formally defined as $\rho_i = \{v\,|\,P_{2v} = i,\,v\in\mathcal{V}\}$ where $P$ is the partition matrix and $\mathcal{V}$ is the set of all nodes. Instead, the subsets of nodes belonging to a module are defined as $\mu_i = \{v\,|\,P_{1v} = i,\,v\in\mathcal{V}\}$ and a module $i$ is said to belong to population $\rho_j$ if $\mu_i\subset\rho_j$. To ease comparisons with other studies and real brain structural connectivities, we also computed the modularity index
\begin{equation}
    Q = \frac{1}{2|\mathcal{E}|}\sum_{c}\left[e_{c} - \left(\frac{\sum_{i\in{c}}k_{i}}{2|\mathcal{E}|}\right)^{2}\right],
\end{equation}
where each $c$ in the summation denotes a community given a partition $P'$, $e_c$ is the number of edges in $c$, $k_i$ is the degree of a node $i\in{c}$, and $|\mathcal{E}|$ is the number of edges of a network $\mathcal{G}=(\mathcal{V},\mathcal{E})$ with $\mathcal{V}$ and $\mathcal{E}\in\mathcal{V}\times\mathcal{V}$ being the set of nodes and edges, respectively. In \hyperref[fig:ExampleMatrices]{Fig. 2D,E}, we report the modularity index $Q_{l}$ of the networks generated with our variation of the nSBM using the partition of nodes in layers $l=1,2$.

\subsection{Dynamical Model and Measures}
We consider the underlying network to be static, and associate each node to a Kuramoto-Sakaguchi oscillator. The phase $\theta_i$ of each oscillator $i$ is governed by the following equation of motion:
\begin{equation}\label{eqn:KSdynamics}
    \frac{\mathrm{d}\theta_i}{\mathrm{d}t} = \omega_i - K\sum_{j}^{N}A_{ij}\sin\left(\theta_j - \theta_i - \alpha_{ij} \right)
\end{equation}
where $A = \{A_{ij}\}_{i,j\in\mathcal{V}}$ is the adjacency matrix, $\omega_i$ is the natural frequency of oscillator $i$, $\alpha_{ij}$ is the phase lag between oscillators $i$ and $j$, and $K$ is the coupling constant. As in the classical models for chimera states and metastability in oscillatory networks \cite{Abrams2008, Shanahan2010, Panaggio2015}, we used identical oscillators by choosing $\omega_i=\omega=1$. We also fixed the phase lag to be equal to zero if $i,j$ are in the same module at layer $l=1$, and $\alpha_{ij} = \alpha$ otherwise. The zero phase lag for within-module connections at the lowest hierarchical layer was chosen as an approximation of shorter phase delays between oscillators in the same module of real networks embedded in a metric space. As standard practice in the dynamical systems' literature \cite{Panaggio2015}, we parametrized the phase lag $\alpha$ using the lag parameter $\beta = \pi / 2 - \alpha$. In this manuscript, we report results for $\beta = 0.1$ which exemplifies our results. Since the oscillators are identical (i.e., they have the same natural frequency), the coupling constant effectively acts as a time re-scaling factor, leaving results qualitatively invariant. Here, numerical simulations were performed using a coupling constant $K=50/k$ normalized by the average degree of the network, allowing us to compare networks of the same size for varying degree. All simulations were performed using the Euler method with steps $\Delta{t}$ of size $0.001$ for a total of $55000$ steps with $5000$ steps of relaxation. We confirmed that this step size was appropriate by verifying that results were unchanged when using smaller Euler step sizes (results not shown).
\newline

\subsubsection{Kuramoto Order Parameter}
To quantify the degree of synchrony of the whole system we used the Kuramoto Order Parameter (KOP)
\begin{equation}\label{eqn:whole_KOP}
    Z = Re^{i\psi} = \frac{1}{N}\sum^{N}_{j}\exp(i\theta_j),
\end{equation}
where the real part $R\in[0,1]$ quantifies the degree of synchronization, with $R\sim1/\sqrt{N}$ suggesting incoherent motion and $R\sim1$ full synchronization.

\subsubsection{Local Kuramoto Order Parameter}
In a similar fashion, we defined the local KOP for a subset $\mu\subset N$ of oscillators as
\begin{equation}
    Z_{\mu} = R_{\mu}e^{i\psi_{\mu}} = \frac{1}{|\mu|}\sum_{j\in \mu}\exp(i\theta_j).
\end{equation}
Similarly to Eq. (\ref{eqn:whole_KOP}), the real part of this measure, $R_\mu\in[0,1]$, quantifies the degree of coherence of the subset of oscillators $\mu$ in isolation.

\subsubsection{Measures for Metastability}
We used the metastability index $\sigma_{\mathrm{met}}(R_{j})$, introduced in \cite{Shanahan2010}, as a measure of metastability. In its original form, $\sigma_{\mathrm{met}}(R_{j})$ is obtained by computing the average standard deviation of the local KOPs $R_{j}$. Since the $3$-layer hierarchical model is equipped with a partition matrix defining distinct blocks in each layer, the index of metastability can be calculated in each layer $l$. Consequently, here we define a distinct index $\sigma^{l}_{\mathrm{met}}$ for each layer $l$. For instance, consider the set of blocks in layer $1$, then the degree of metastability of the layer $1$ blocks is $\sigma^{1}_{\mathrm{met}} = \langle\sigma(R_{\mu_i}(t))\rangle_{i\in[1, B_1]}$, where $B_1$ is the number of blocks in the first layer, $\mu_i$ with $i\in[1, B_1]$ identifies the set of oscillators that make up a block $i$ in the first layer, and $\sigma$ is the standard deviation. Similarly, we define $\sigma^{2}_{\mathrm{met}}$ for layer $2$ blocks. For layer $3$ instead, the metastability index is just the variation of the whole system KOP since all oscillators belong to the same block, i.e., $\sigma^{3}_{\text{met}} = \sigma(R)$.

\subsubsection{Measures for Chimera States}
The term chimera state is generally invoked to denote the coexistence of coherent and incoherent patterns in systems of identical oscillators. Over the past decade, various measures have been defined to characterize different types of chimera states \cite{Kemeth2016, Haugland2021}. In the case of identical sinusoidally-coupled oscillators with fixed amplitude, the local KOPs can be used to measure the degree of phase-coherence \cite{Panaggio2015} between the system's parts. However, since an analytical characterization of chimera states is only possible in the limit of an infinite number of oscillators, an arbitrary threshold needs to be introduced, $\delta_1$, to numerically assess the presence of chimera states \cite{Kemeth2016}. For instance, if a system self-organizes into two subsets $\rho_1, \rho_2$ of oscillators with different degrees of synchronization at some time $t$, i.e., $d(t) = |R_{\rho_1}(t) - R_{\rho_2}(t)| > \delta_1$, we conclude that system is in a chimera state. Here, due to the similarity with the classic two-population model \cite{Abrams2008}, we focus on chimera states in layer $2$. Specifically, we numerically characterized chimera states by computing the mean and standard deviation of the difference between the populations' local KOPs in time:
\begin{eqnarray}
    \bar{d} &=& \frac{1}{T-r}\sum_{t=r+1}^{T}d(t) \label{eqn:average_difference}\\
    \sigma(d) &=& \sqrt{\frac{1}{T-r}\sum_{t=r+1}^{T}\left(d(t)-\bar{d}\,\right)},
\end{eqnarray}
where $T$ is the number of steps in the simulation and $r$ the number of relaxation steps. These measures were then averaged over $100$ distinct initial conditions. Using these measures and an additional threshold $\delta_2$, we can distinguish three types of chimera states:
\begin{itemize}
    \item stable chimera state \cite{Panaggio2015}: a chimera state in which the local KOPs are different and remain constant, i.e., $\bar{d} > \delta_1$ and $\sigma(d) < \delta_2$;
    \item breathing chimera state \cite{Panaggio2015}: a chimera state in which one of the local KOPs oscillates, i.e., $\bar{d} > \delta_1$ and $\sigma(d) > \delta_2$;
    \item metastable and alternating chimera state \cite{Shanahan2010, Buscarino2015, Feng2023}: the state of a system in which the degree of synchronization of all subsets of oscillators varies irregularly or in an alternate manner, i.e., $\bar{d} < \delta_1$ and $\sigma(d) > \delta_2$;
\end{itemize}
where thresholds $\delta_1$ and $\delta_2$ are chosen to be $3$ standard deviations higher than the baseline mean $\bar{d}$ and standard deviation $\sigma(d)$ calculated in the case of no mesoscale structure, i.e., $H=0$ (see also sensitivity analysis in \hyperref[sec:appendixB]{Appendix B}, \hyperref[sec:fig_sensitivity]{Fig. 10}). These measures allow us to identify the regions of the parameter space in which symmetry breaking states occur in the second layer. This classification scheme is similar to that described in \cite{Kemeth2016}, with the main difference being the correlation measure. While the local curvature measure used in \cite{Kemeth2016} is a more general correlation measure, its application to our case is not practical as it would require approximating the phase of a modules or population using the local order parameters to detect symmetry breaking states in a specific layer.
\newline

\subsection{Data Availability}
All code and notebooks related to this work can be downloaded from \cite{Caprioglio2024}.

\section{Results}\label{sec:results}
\subsection{Metastable dynamics are detected at the whole-system level}\label{sec:Results0}
\begin{figure}[h!]
    \centering
    \includegraphics[width=1.0\textwidth]{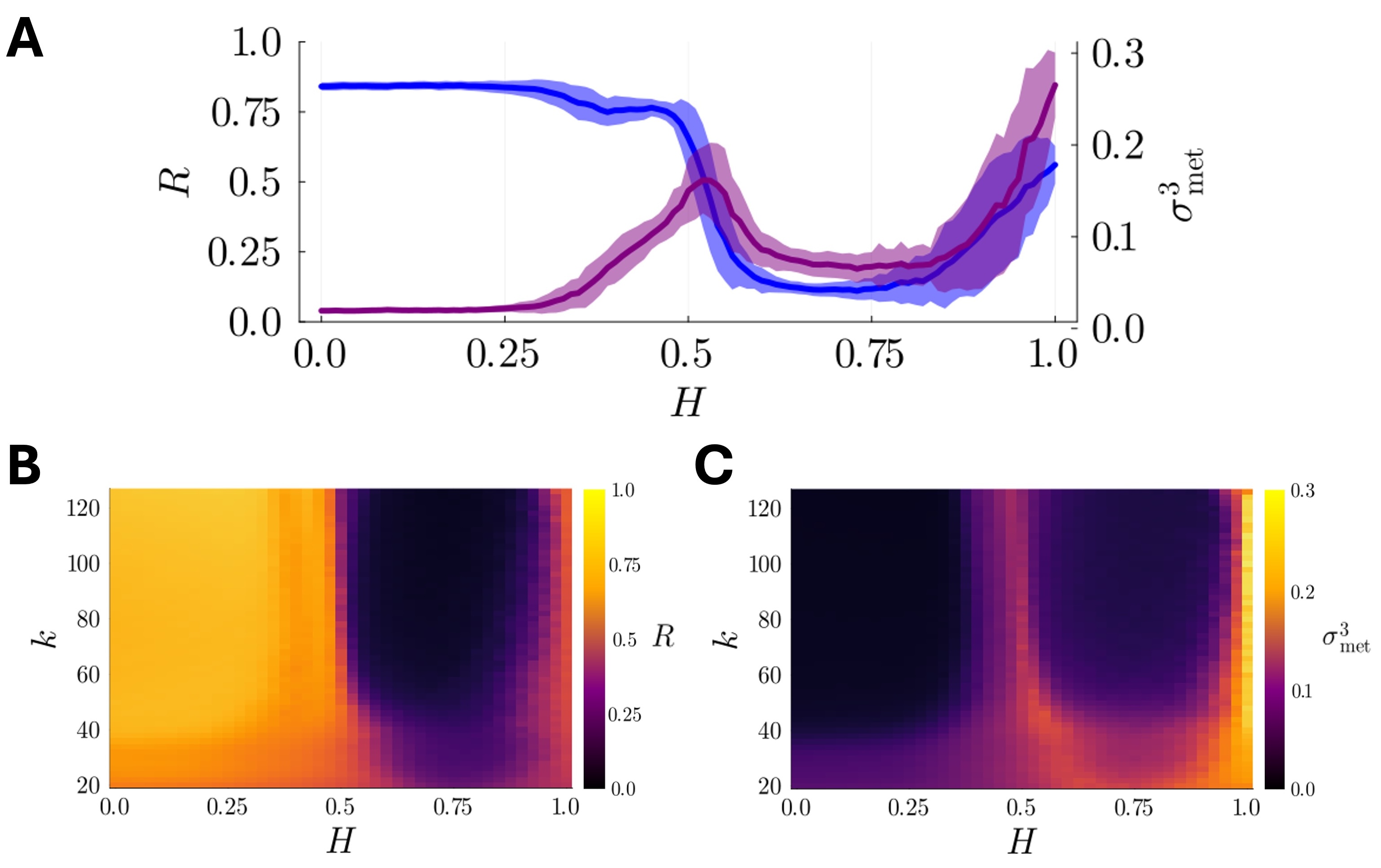}
    \caption{(\textbf{A}) Degree of synchronization, $R$, and index of metastability, $\sigma^{3}_{\text{met}}$, in the $l=3$ (whole-system) layer as a function of the structural parameter $H$. (\textbf{B}) Degree of synchronization $R$ as a function of $H$ and $k$. (\textbf{C}) Index of metastability $\sigma^{3}_{\text{met}}$ as a function of $H$ and $k$. Results were obtained using $n_1 = 16$, $n_2 = 8$, and $k = 51.2$, averaged over 100 seeds.}
    \label{fig:Results0}
\end{figure}
To detect metastable dynamics and chimera states, we numerically analyzed the dynamics of Kuramoto-Sakaguchi oscillators on $3$-layer hierarchical networks at different levels of observation. We started by examining the dynamics at the whole-system level, presenting the KOP and $\sigma^{3}_{\text{met}}$ for networks of size $N = 512$ with $n_1 = 16, n_2 = 8$ in \hyperref[fig:Results0]{Fig. 3}. Our analysis of the whole parameter space (\hyperref[fig:Results0]{Fig. 3B,C}) suggests that, while there is some degree of variability depending on the average degree of the network, $k$, the degree of synchronization of the system and the index of metastability mostly depend on the structural parameter $H$. Thus, we analysed in more detail our results for fixed average degree $k=N/5=51.2$, which exemplifies the results obtained for both smaller and larger average degrees (\hyperref[fig:Results0]{Fig. 3A}). Metastable dynamics are detected only when mesoscale structures are well defined ($H>0.3$, see also \hyperref[fig:ExampleMatrices]{Fig. 2D}) in good agreement with our hypothesis. In contrast, in the absence of well-defined populations (i.e., low values of $H$) the system displays high coherence and a low index of metastability. As the connectivity within populations increases further, metastability reaches a local maximum around $H=0.5$ while the whole-system degree of synchronization decreases sharply. Beyond $H>0.7$, the density of connections between populations is very low (see \hyperref[fig:ExampleMatrices]{Fig. 2D}). Consequently, the metastable index increases sharply while the KOP approaches $0.5$, indicating that the system has decoupled into two independent populations. While metastable dynamics can already be detected at the whole-system level in the presence of well-defined populations, global measures do not provide insights into why metastability starts to increase only for values of $H>0.3$.

\subsection{Self-organized chimera states are detected at the population layer}\label{sec:results1}
\begin{figure}[h!]
    \centering
    \includegraphics[width=0.95\textwidth]{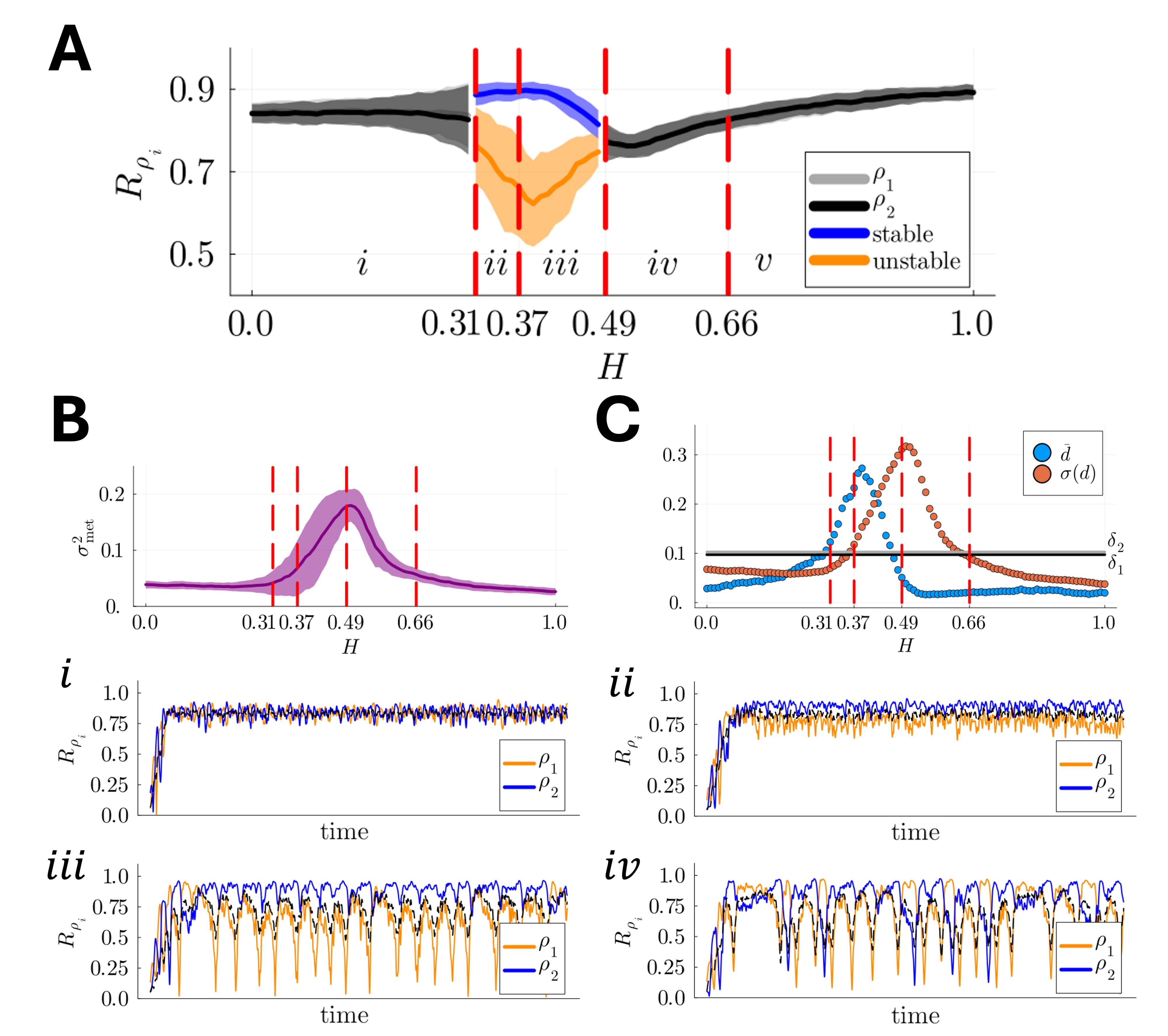}
    \caption{(\textbf{A}) Mean degree of synchronization $R_{\rho_i},\,i=1,2$ for varying values of $H$. (\textbf{B}) Populations' metastability index, $\sigma^{2}_{\mathrm{met}}$, as a function of $H$. (\textbf{C}) (Blue line) mean difference between the populations' local KOP $\bar{d}$ and (orange line) standard deviation of the difference between local KOPs, $\sigma(d)$ as a function of $H$.
    Red dashed lines separate the regions of the parameter space in which symmetry-breaking states are detected. ($i$) For low values of $H$ the system displays coherent dynamics and the two populations display the same degree of synchronization. ($ii, iii$) For intermediate values of $H$ stable and breathing chimera states emerge, respectively.  ($iv$) Metastable chimera states emerge for $H=0.5$. Results obtained using $n_1 = 16$, $n_2 = 8$, and $k = 51.2$, averaged over 100 seeds.}
    \label{fig:Results1}
\end{figure}

To understand why metastable dynamics are detected as the connectivity within population increases, we analysed the populations' local KOPs, $R_{\rho_i}$, and the populations' metastability index, $\sigma^{2}_{\mathrm{met}}$ for varying values of $H$. We computed the values of $\bar{d}$ and $\sigma(d)$ to identify chimera states (\hyperref[fig:Results1]{Fig. 4C}) as illustrated in the Methods section. Independently of the value of $k$, we identified $5$ regions of the parameter space, $i-v$, in which distinct states are detected in the population layer. Symmetry breaking states are detected in regions $ii-iv$, with stable ($ii$), breathing ($iii$), metastable and alternating ($iv$) chimera states. When stable and breathing chimera states are detected, we define as stable population the subset of oscillators displaying the higher mean local KOP, and as unstable population the subset of oscillators with a lower degree of synchronization. As an example, in \hyperref[fig:Results1]{Fig. 4($iii$)}, we identified $\rho_2$ as the stable population and $\rho_1$ as the unstable (breathing in this case) population. In \hyperref[fig:Results1]{Fig. 4}, we present our results for $k=51.2$.
\newline

In good agreement with our results of Sec. \ref{sec:Results0}, metastability was found to reach its minimum in region $i$, where the absolute difference between local KOPs ($\bar{d}$ from Eq.(\ref{eqn:average_difference})) was below the threshold $\delta_1$. This confirms that the system reaches a symmetric coherent global state (for instance, see \hyperref[fig:Results1]{Fig. 3($i$)}) in the absence of well-defined mesoscale structures. In this region, both populations display a high degree of synchronization, equal to that of the whole system, $R = 0.84 \pm 0.04$, with some fluctuations due to phase lag frustration.
\newline

In the same region of the parameter space in which metastable dynamics are initially detected at the whole-system layer, we encounter stable and breathing chimera states (regions $ii$ and $iii$) in the second layer. Within these regions, the system self-organizes into a symmetry breaking state in which one of the two populations is more synchronized than the other, such that $\bar{d}>\delta_1$. While in region $ii$ the unstable population displays a stable degree of synchronization ($\sigma(d)<\delta_2$), in region $iii$ the unstable population's local KOP presents oscillatory dynamics ($\sigma(d)>\delta_2$). This result suggests that the initial increase in metastability $\sigma^{3}_{\mathrm{met}}$ detected at the whole-system level reflects the presence of stable and breathing chimeras in the second layer.
\newline

Metastable and alternating chimeras are detected in the range $0.49<H<0.66$ in which $\bar{d}<\delta_1$ and $\sigma(d)>\delta_2$. In the vicinity of $H=0.5$, both local KOPs display a similar degree of synchronization and the populations' metastability, $\sigma^{2}_{\text{met}}$, is maximised. When the metastability index is at its maximum, populations (de)synchronize in a non-predictable manner. For instance, in \hyperref[fig:Results1]{Fig. 4($iii$)}, when a population desynchronizes the other populations' response can either be an increase or decrease in synchronization. In the same region, as the parameter $H$ is increased further, the number of connections between the two populations decreases and alternating chimera states can also found (see \hyperref[sec:appendixB]{Appendix B} \hyperref[fig:alternating]{Fig. 12}).
\newline

In region $v$ both populations display a high local KOP but they are not synchronized between each other, indicating once again that beyond $H \sim 0.7$ the two populations become effectively uncoupled. In \hyperref[fig:Results1]{Fig. 4($i-iv$)}, we show examples of the evolution of the local KOPs over time for four exemplifying values of the structural parameter $H$.

\subsection{Chimera states can be supported by metastable modules}\label{sec:Results_layer1}
The remaining question is whether modules in the first layer also exhibit metastable dynamics, and whether modules in different populations display distinct dynamics depending on the symmetry-breaking states detected in layer $2$.
\newline

\begin{figure}[h!]
    \centering
    \includegraphics[width=1\textwidth]{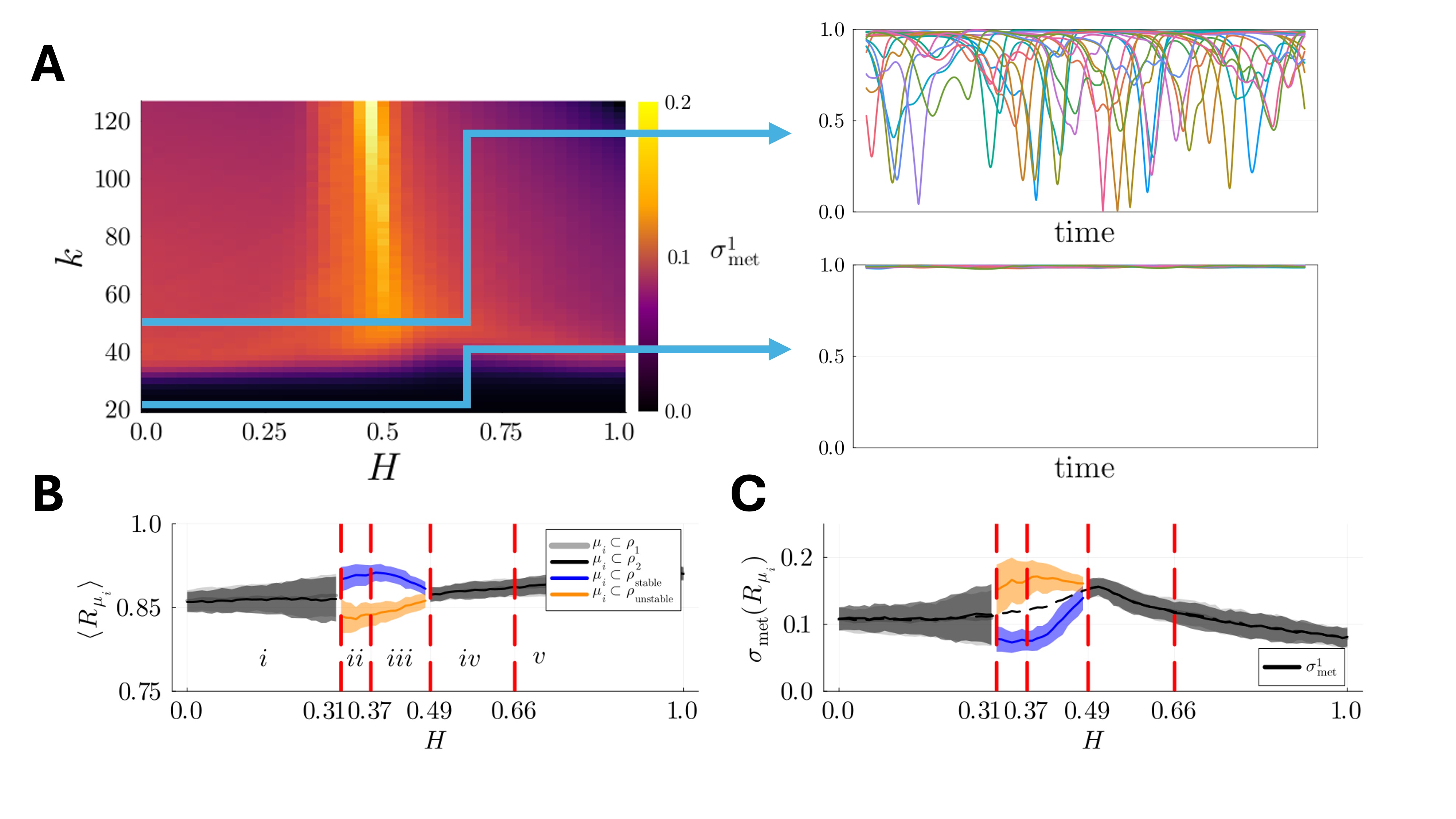}
    \caption{(\textbf{A}) Left: Metastability index $\sigma^1_{\text{met}}$ calculated over all modules $\mu_{i}$ as a function of $H$ and $k$. Right: example of $R_{\mu_i}$ dynamics in time for a single simulation with average degree $k=51.2$ (top) and $k=21$ (bottom). (\textbf{B}) Average local KOP within each population $j=1,2$, i.e., $\langle R_{\mu_i}\rangle$ for $\;\mu_i\subset\rho_j$, for fixed $k=51.2$ and varying values of $H$. (\textbf{C}) Metastability index within each population $j=1,2$, i.e., $\sigma_{\mathrm{met}}(R_{\mu_i}),$ for $\;\mu_i\subset\rho_j$, for fixed $k=51.2$ and varying values of $H$, and overall metastability of the first layer blocks $\sigma^1_{\mathrm{met}}$ (black dashed line). Red dashed lines separate the regions of the parameter space in which symmetry-breaking states are detected in layer $2$.}
    \label{fig:results2}
\end{figure}

To answer the first question, we computed the degree of metastability $\sigma^{1}_{\mathrm{met}}$ for varying values of $H$ and $k$ (\hyperref[fig:Results2]{Fig. 5A}). While in layers $2$ and $3$., metastable dynamics were dependent on the structural parameter $H$, our analysis shows that in layer $1$ the dominant parameter is the average degree of the network $k$. For systems of size $n_1=16$, $n_2=8$, the modules' degree of metastability was zero for values of $k$ below $\sim 30$, the same rage of values in which modularity in layer $1$ is the highest (\hyperref[fig:ExampleMatrices]{Fig. 2E}). In this regime, all modules remain fully synchronized for all values of $H$ (see \hyperref[fig:Results2]{Fig. 5A$(a)$} for an example). As $k$ increases and modularity in layer $1$ decreases, modules start to display metastable dynamics modulated by the value of $H$ (see \hyperref[fig:Results2]{Fig. 5A$(b)$}). Likewise, as in the previous subsection, for fixed $k>30$, the metastability index calculated for all modules peaks at $H=0.5$. However, as $k$ is increased further, the metastability index slightly increases only around $H=0.5$ while slightly decreasing for all other values of $H$ (see \hyperref[fig:metastability_layer_1_var_k]{Fig. 13} in \hyperref[sec:appendixB]{Appendix B}). Thus, our analysis indicates that while the average connectivity of the network and modularity in layer $1$ determines whether spontaneous (de)synchronization manifests or not in modules in layer $1$, the presence of a mesoscale structure modulates the degree of metastability.
\newline

To address our second question, we computed the average local KOP and the degree of metastability of modules within each population separately, with fixed $k=51.2$ and varying values of $H$ (\hyperref[fig:Results2]{Fig. 5B,C}). Here, the metastability index of modules in population $\rho_j$ was defined as $\sigma_{\text{met}}(R_{\mu_{i}})=\langle\sigma(R_{\mu_{i}})\rangle_{\mu_{i}\subset\rho_j}$. We divided the parameter range into the same regions $i-v$ as presented in the previous section (see \hyperref[fig:Results1]{Fig. 4C}). In the absence of a well-defined mesoscale structure (region $i$), modules in each population display the same index of metastability and the same average degree of synchronization. In contrast, in regions $ii$ and $iii$, modules in the stable and unstable populations display different degrees of synchronization and metastability. The stable population displays a higher average KOP and lower metastability while the modules in the unstable population exhibit the opposite. As expected, in regions $iv$ and $v$, modules in both populations display the same index of metastability, as well as the same degree of synchronization. Therefore, it is only in the presence of stable and breathing chimeras in  layer $2$ that modules in distinct populations display different degrees of metastability, reflecting the stability or instability of the populations to which the modules belong.

\subsection{Relationship between symmetry-breaking states and characteristic timescales of the system}\label{sec:Results_Laplacian}
In the previous sections, we  provided numerical evidence for the emergence of chimera and metastable states in hierarchically modular networks at different levels of observation. While in layer $2$ different types of chimera states were identified depending on the mesoscale structure of the network (determined by $H$), in layer $1$ we showed how metastability arises as the average connectivity of the network reaches a certain threshold. In this section, we aim to explain these observed phenomena by analyzing the spectral information of the graph Laplacian.
\newline

\begin{figure}[h!]
    \centering
    \includegraphics[width=0.9\textwidth]{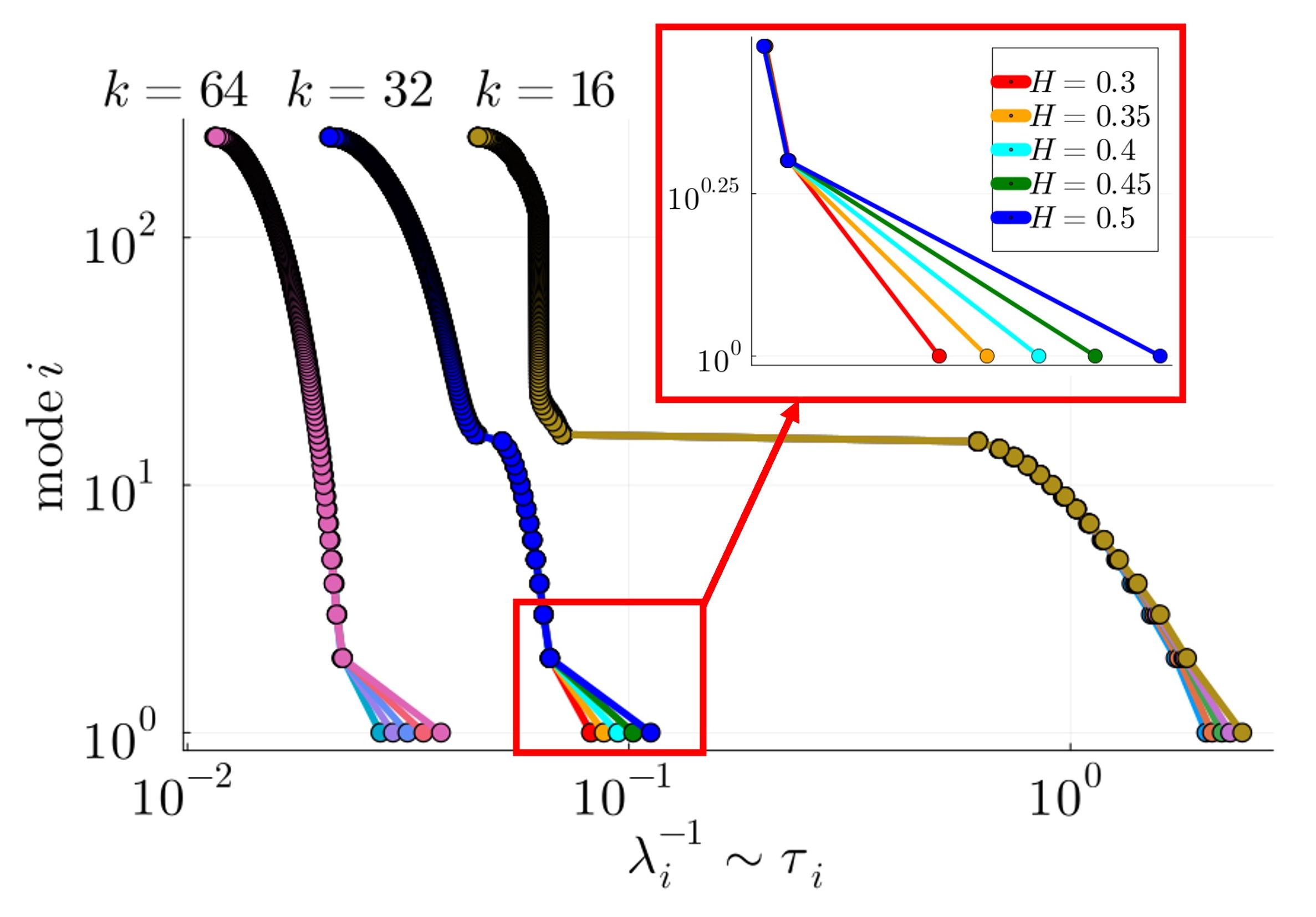}
    \caption{Gaps between consecutive eigenvalues $\lambda_{2}, \lambda_{3},\,\dots\,\lambda_{N}$ are affected by the structural parameter $H$ and the average degree $k$. In this model, increasing $k$ reduces the size of the second spectral gap, between $\lambda_{B_{1}}$ and $\lambda_{B_{1}+1}$ (where $B_1=16$ is the number of layer $1$ blocks). On the other hand, increasing $H$ increases the size of the first spectral gap (inset figure), between $\lambda_{B_{2}}$ and $\lambda_{B_{2}+1}$ (where $B_2=2$ is the number of layer $2$ blocks). Each eigenvalue was obtained by averaging over $10$ randomly generated $3$-layer networks with $n_1=16, n_2=8$.}
    \label{fig:Results3_Part1}
\end{figure}

The systems we investigated are structurally homogeneous in each layer (uniform degree distribution). Nonetheless, a peculiar property of diffusive-like dynamics on hierarchically modular networks, such as synchronization dynamics, is the richness of distinct information-diffusion pathways across different timescales \cite{Villegas2022}. In the absence of phase frustration, blocks of oscillators synchronize hierarchically, from layer $1$,until the whole system is synchronized \cite{Arenas2006}. As shown in \cite{Arenas2006}, the linearized Kuramoto model (without phase lags) reduces to a linear diffusion model governed by the graph Laplacian. The graph Laplacian is defined as $L = D-A$ with eigenvalues $\lambda_i$ ranked as $0 = \lambda_1 < \lambda_2 <\,\dots\,<\lambda_{N}$, where $A$ is the adjacency matrix of the network and $D$ is a diagonal matrix with $D_{ii} = \sum_{j}^{N}A_{ij}$. Crucially, authors in \cite{Arenas2006} showed how the gaps between eigenvalues, which separate blocks of nodes at distinct hierarchical layers, are associated with the relative difference in characteristic timescales between hierarchical layers. When the spectral gaps are small, the relative difference between hierarchical timescales is also small, and blocks at distinct hierarchical layers are not well defined. Conversely, when the spectral gaps are large, the opposite is true. Owing to the similarities with the networks studied in \cite{Arenas2006}, our Laplacian spectrum analysis (depicted in \hyperref[fig:Results3_Part1]{Fig. 6}) also reveals similar spectral gaps depending on the value of $k$ and $H$. In particular, we identify two spectral gaps:
\begin{itemize}
    \item the first spectral gap is found to be between $\lambda^{-1}_{B_2}$ and $\lambda^{-1}_{B_2+1}$, separating the slowest mode associated with the characteristic timescales of layer $3$ and the slow modes associated with layer $2$,
    \item the second spectral gap is found to be between $\lambda^{-1}_{B_1}$ and $\lambda^{-1}_{B_1+1}$, separating the slow modes associated with the characteristic timescales of layer $2$ and the fast modes associated with layer $1$,
\end{itemize}
where $B_1$ and $B_2$ are the number of blocks in layers $1$ and $2$, respectively, and the inverse of each eigenvalue $\lambda_i$ is indicative of the characteristic timescale of mode $i$, i.e., $\lambda_{i}^{-1}\sim\tau_i$. Informed by these observations, we hypothesised that (a) the eigenmodes associated with the slower timescales of the system, $\lambda_{j}$ for $j\in[1,B_1]$, determine the state of the system in layers $2$ and $3$, and that (b) the second spectral gap between fast and slow eigenmodes affects the synchronizability and metastable dynamics of modules in layer $1$.
\newline

\begin{figure}[h!]
    \centering
    \includegraphics[width=1.0\textwidth]{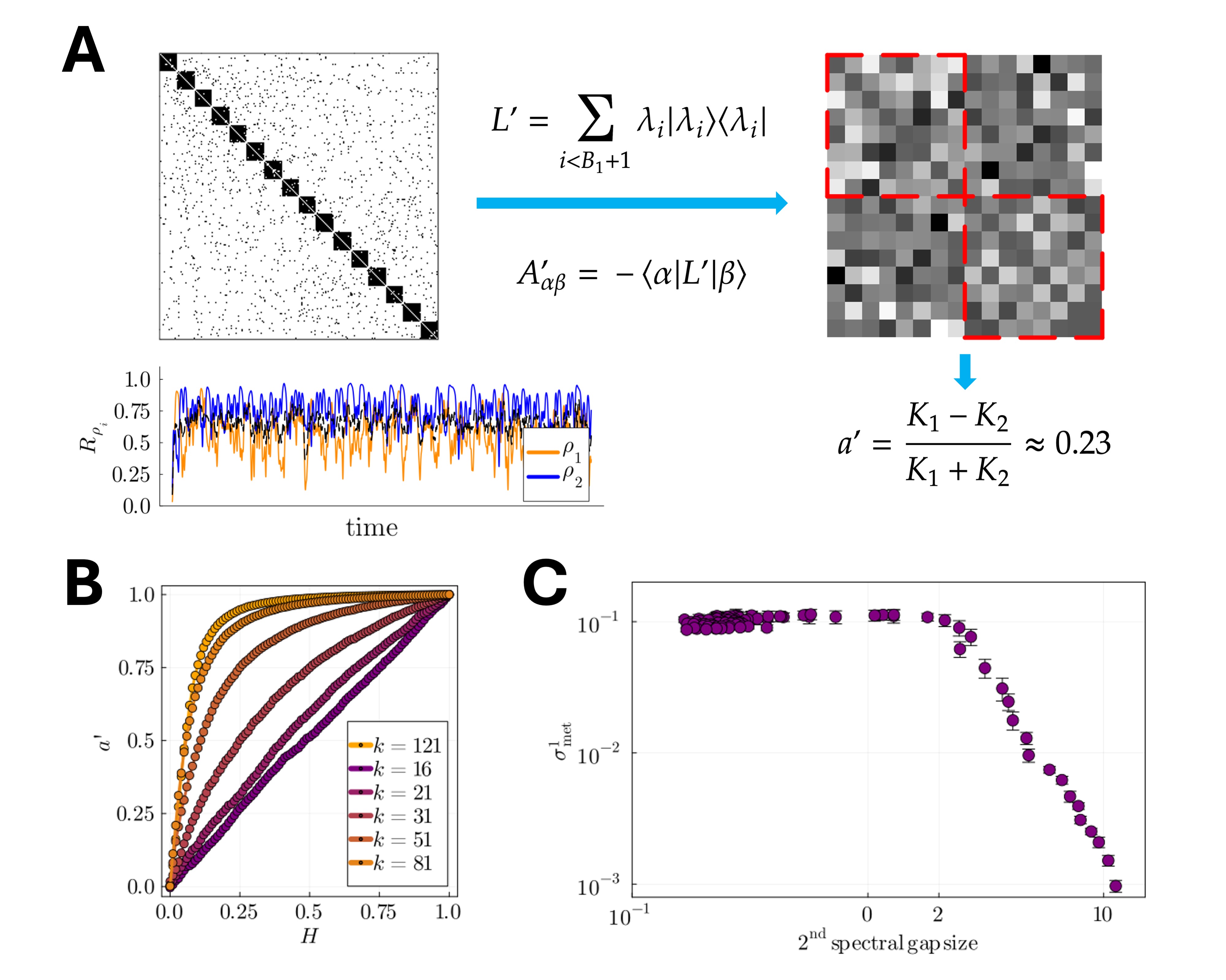}
    \caption{(A) Left: example of populations' dynamics in a stable chimera state in layer $2$, where $A$ was obtained using the $3$-layer variation of the nSBM with $H=0.2$, $n_1=16,\,n_2=8$, and $k=21$. Right: the weighted coarse-grained adjacency matrix $A'$ is obtained by time-rescaling the Laplacian associated with the original adjacency matrix $A$. The symmetry breaking parameter $a'=0.23$, obtained by computing the normalized difference between intra-population $K_1$ (red squares) and inter-population $K_2$ couplings of the coarse-grained adjacency matrix $A'$, is also associated with stable chimera states in the two-population model (see \cite{Abrams2008} and results therein). (B) The symmetry breaking parameter $a'$ for varying $H$ and six exemplifying average degrees $k$. (C) Log-log plot displaying the cutoff point when metastability in layer $1$, $\sigma^{1}_{\mathrm{met}}$, starts to sharply decrease as the size of the second spectral gap, $\lambda_{B_1+1} - \lambda_{B_1}$, increases.}
    \label{fig:Results3_Part2}
\end{figure}

To test hypothesis (a), we adopted the Laplacian Renormalization Group (LRG) flow approach recently introduced in \cite{Villegas2022, Villegas2023}. In summary, this approach allows us to rescale the system up to a timescale $1/\lambda^{*} \sim \tau^{*}$ by discarding the fast modes associated with eigenvalues $\lambda_i > \lambda^{*}$. Then, the resulting coarse grained structure is obtained by using the slow modes associated with $\lambda_i < \lambda^{*}$ only. We adopt the bra-ket notation used in \cite{Villegas2023}, such that $|\lambda_i\rangle$ denotes the column eigenvector associated with eigenvalue $\lambda_i$ while $\langle\lambda_i|$ is its corresponding row vector. Then, the dot product can be written as $\langle\lambda_j|\lambda_i\rangle$ while the outer product is indicated by $|\lambda_i\rangle\langle\lambda_j|$. Using this notation, we rewrote the graph Laplacian in terms of its eigenmodes $|\lambda_{i}\rangle$, i.e.,
\begin{equation}
    L = \sum_{i}^{N}\lambda_{i}|\lambda_{i}\rangle \langle\lambda_{i}|,
\end{equation}
and constructed the time-rescaled Laplacian at $\tau^{*}=1/\lambda^{*}=1/\lambda_{B_{1}+1}$, as
\begin{equation}
    L' = \sum_{i<B_{1}+1}^{N}\lambda_{i}|\lambda_{i}\rangle \langle\lambda_{i}|,
\end{equation}
discarding the $N-B_{1}$ fast modes of the system. Next, we constructed the adjacency matrix $A'$ from the resulting time-rescaled Laplacian $L'$. This was done by aggregating nodes within each of the $B_{1}$ modules. Specifically, we defined $B_{1}$ orthogonal vectors $|\alpha\rangle = (\alpha_{1}, \alpha_{2},\,\dots\,\alpha_{N})'$, $\alpha = 1, 2,\,\dots\,B_{1}$, using the partition of the $3$-layer hierarchical networks at layer $1$, i.e., $\alpha_{i}=1$ if $P_{1\alpha}=\alpha$ and $\alpha_{i}=0$ otherwise. Each element of the $B_{1}\times B_{1}$ adjacency matrix $A'$ was obtained by projecting the rescaled Laplacian onto the new $B_{1}$-dimensional basis defined by the vectors $|\alpha\rangle$, i.e., $A'_{\alpha\beta} = - \langle\alpha|L'|\beta\rangle$, with $A'_{\alpha\alpha}=0$ to avoid self interactions. The resulting weighted fully-connected adjacency matrix $A'$ (an example is shown in \hyperref[fig:Results3_Part2]{Fig. 7A}) shares similar properties with the weighted adjacency matrix of the analytically solvable two-population model. Specifically, pairs of nodes in the same population are more strongly connected than pairs of nodes in distinct populations.
\newline

While $A'$ loses the fine-grained structural information in layer $1$, it preserves the structural information in layer $2$ which depends on the structural parameter $H$. This allows a numerical comparison of $a$, the symmetry breaking parameter of the two-population model, with the structural parameter $H$, where $a$ is defined as the difference between the interaction strength between pairs of nodes in the same population and that of between pairs of nodes in different populations. From the reconstructed adjacency matrix $A'$, we obtained a numerical approximation of the symmetry breaking parameter $a'$ by computing the normalized difference between the average strength of interaction between nodes in the same population, $K_1$, and nodes in distinct populations, $K_2$, as illustrated in \hyperref[fig:Results3_Part2]{Fig. 7A}. Next, we investigated the relationship between $a'$ and $H$ for different average degrees of the network (see \hyperref[fig:Results3_Part2]{Fig. 7B}). Our results suggest that there is a linear relationship between $a'$ and $H$ when the average degree is low, and a nonlinear relationship as $k$ increases. Notably, we found that when the relationship is linear or approximately linear, the symmetry-breaking parameter $a'$ predicts well the range of $H$ in which stable and breathing chimeras are detected (for instance, see the case for $k=21$ in \hyperref[fig:Results3_Part2]{Fig. 7A} and \hyperref[sec:appendixB]{Appendix B} \hyperref[fig:identify_chimeras_k_21]{Fig. 11}). However, we found this not to be the case for larger average degrees, such as $k=51.2$ used in the previous sections, when the relationship between $a'$ and $H$ was found to be nonlinear. For instance, using $k=51.2$ we found stable and chimera states in the range $0.31< H < 0.49$ of the structural parameter, which corresponds to the range $0.8<a'<0.9$ of the numerical approximation of the symmetry breaking term, which is well outside the range of values in which stable and breathing chimeras appear in the two-population model \cite{Abrams2008}. This result suggests that, as the second spectral gap decreases (equivalently, as the average degree of the network increases), the slow modes of the graph Laplacian alone do not describe well the behaviours we observe in layer $2$ and that the fast modes contributions should be taken into account.
\newline

To test hypothesis (b), we computed the size of the second spectral gap, $\lambda_{B_1+1} - \lambda_{B_1}$, and the degree of metastability $\sigma^{1}_{\mathrm{met}}$ of layer $1$ modules for varying $k$. As depicted in the log-log plot in \hyperref[fig:Results3_Part2]{Fig. 7A}, the degree of metastability $\sigma^{1}_{\mathrm{met}}$ decreases sharply for values of the spectral gap $\lambda_{B_1+1}-\lambda_{B_1}>2$, indicating a clear cutoff point in the vicinity of $\lambda_{B_1+1}-\lambda_{B_1}\approx2$ for systems of size $n_1=16,\,n_2=8$ with fixed value of $H=0.5$. The same behaviour was found for any other value of $H$. This result suggests that when the timescale separation between hierarchical layers $1$ and $2$ is large enough and modules are well defined, modules of oscillators remain completely synchronized, such that each subset effectively behaves as a single macro-oscillator. In contrast, when there is no clear timescale separation between hierarchical layers $1$ and $2$, the slow modes of the system (associated with $\lambda_i$, $i<B_1$) affect the synchronization dynamics within modules, resulting in random (de)synchronization patterns that persist in the system.

\subsection{Robustness against structural perturbations and heterogeneous natural frequencies}\label{sec:Results_robustness}
\begin{figure}[h!]
    \centering
    \includegraphics[width=0.99\textwidth]{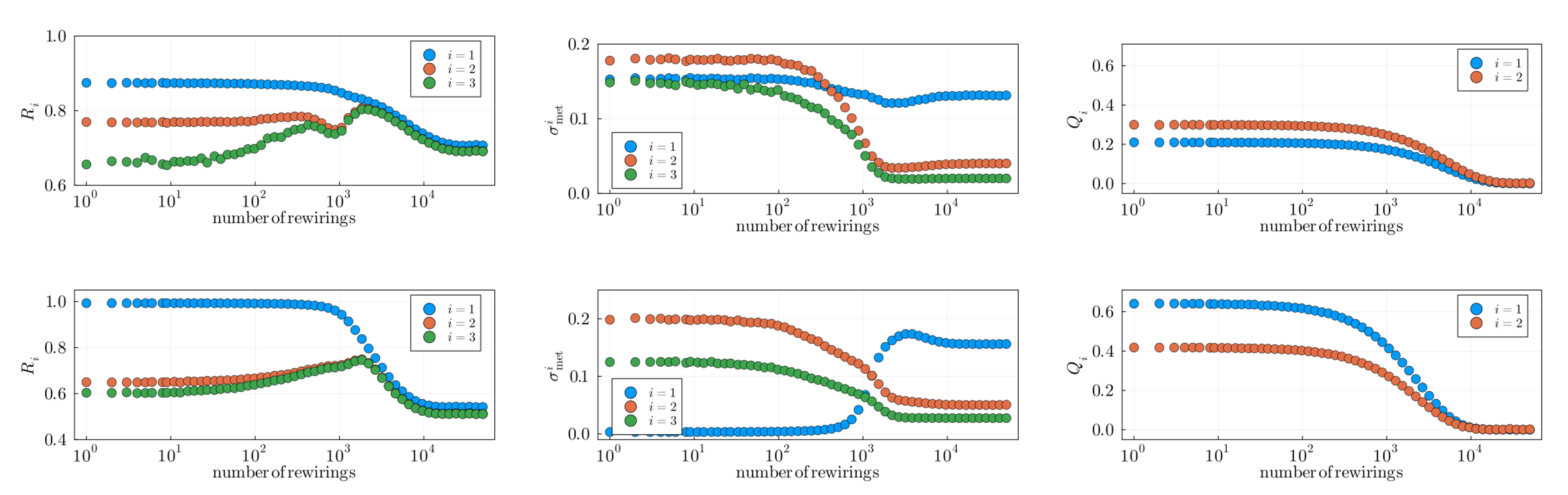}
    \caption{Structural perturbation analysis for fixed $H=0.5$ and for $k=51.2$ (top row) and $k=21$ (bottom row). In each layer $i$, we compute the average degree of synchronization $R_i$ of all blocks in the layer (left column) and the metastability index $\sigma^{i}_{\mathrm{met}}$ (middle column). The modularity index $Q_i$ is computed with respect to the partitions in layer $1$ and $2$ (right column). Results were obtained for systems of size $n_1=16,\,n_2=8$ and averaged over $100$ seeds.}
    \label{fig:Results4}
\end{figure}

\begin{figure}[h!]
    \centering
    \includegraphics[width=0.99\textwidth]{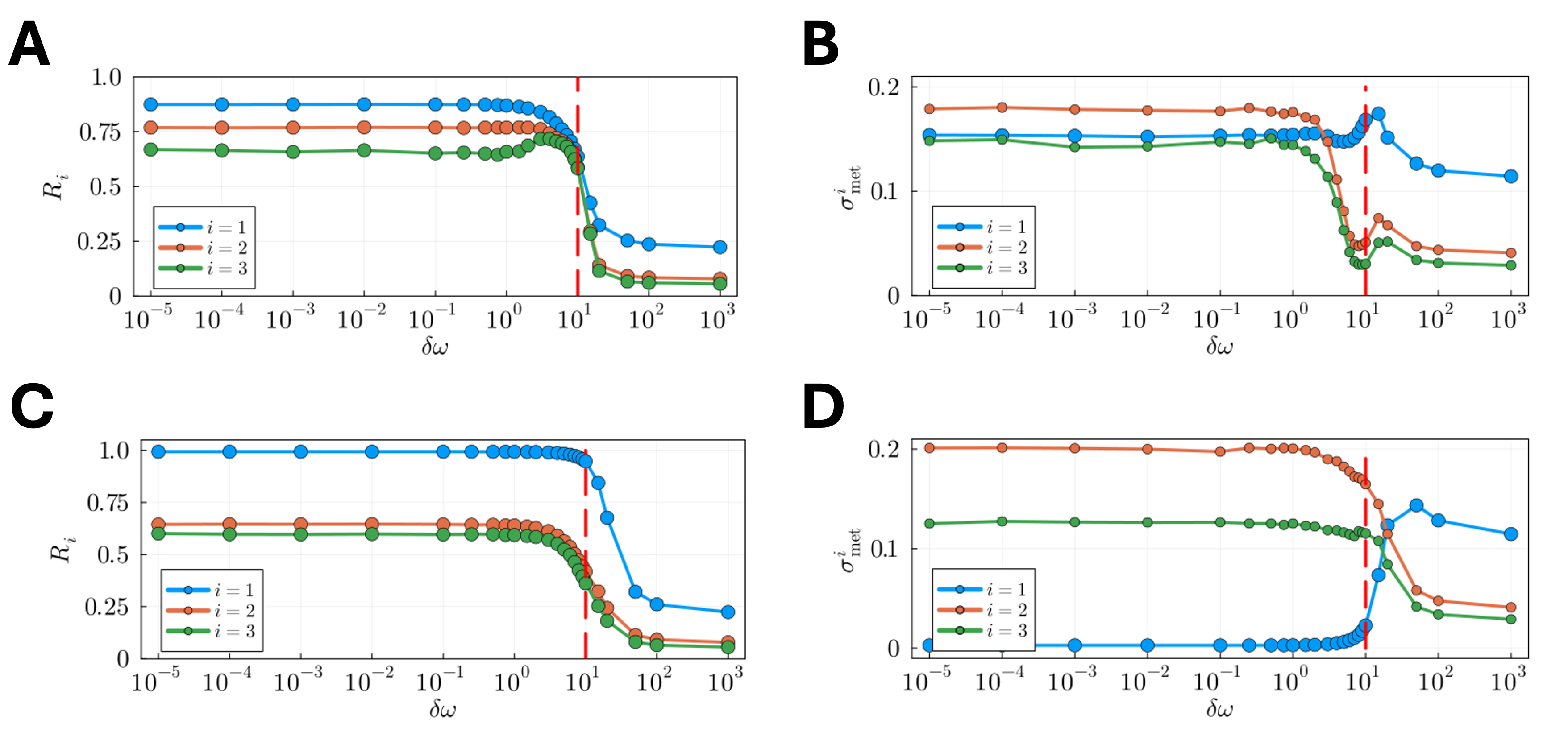}
    \caption{Impact of heterogeneous frequencies with increasing $\delta\omega$ for $H=0.5$ and for $k=51.2$ (top row) and $k=21$ (bottom row). In each layer $i$, we compute the average degree of synchronization $R_i$ of all blocks in layer $i$ (left column) and the metastability index $\sigma^{i}_{\mathrm{met}}$ (right column). Results were obtained for systems of size $n_1=16,\,n_2=8$, using normalized coupling $K/k$ with $K=50$, and averaged over $100$ seeds.}
    \label{fig:Results4_part2}
\end{figure}

Our numerical analyses have shown the relationship between different symmetry-breaking states and the system's hierarchically modular structure. In layers $2$ and $3$, chimera and metastable states are detected as the populations become more defined (with increasing $H$). In layer $1$ instead, we found the opposite relationship: as modules become less defined for increasing $k$ (as shown by decreasing spectral gap and modularity) metastable dynamics are detected. To measure the robustness of these observed dynamics, we analysed systems generated with fixed values of $H$ and $k$ when perturbed in two separate ways: by randomly rewiring $r$ edges in the network (\hyperref[fig:Results4]{Fig. 8}), and by including heterogeneous frequencies $\omega_i$ drawn from a normal distribution $\mathcal{N}(1, \delta\omega)$ with mean $1$ and increasing standard deviation $\delta\omega$ (\hyperref[fig:Results4_part2]{Fig. 9}). We report results for systems constructed using $H=0.5$, since most peaks in metastability in all layers were found for this value, and using two exemplifying average degrees $k=51.2$ and $k=21$, since metastability in layer $1$ was found to depend on the size of the second spectral gap, itself controlled by the average degree of the network.

By randomly rewiring $r$ edges in the network, the average degree of the network remains fixed while modularity in layers $2$ and $1$ is progressively lost (\hyperref[fig:Results4]{Fig. 4}). We found that the system starts to lose its modular features when $\sim10\%$ of the edges are rewired, i.e., $r=10^3$ since the expected number of edges is $\mathcal{E} = kN$, and completely loses modularity (as well as the spectral gaps) when $10^4$ edges are rewired. As expected, measures of metastability and synchrony behave differently in layer $1$ compared to layers $2$ and $3$. While the system loses its metastable features in layers $2$ and $3$ when $\sim10\%$ of the edges are rewired, in layer $1$ metastability slightly decreases in the case of $k=51.2$ and sharply increases in the case of $k=21$. This analysis reflect the fact that the presence of a mesoscale structure in layer $2$ and $3$ is necessary for metastable dynamics to emerge in the upper layers. Modules of oscillators in layer $1$, instead, are characterized by the absence of phase lags between pairs of oscillators in the same module, and therefore display metastable dynamics only when their modular structure is less defined, or, equivalently, when the second spectral gap disappears (which is the case for $k=51.2$). While not the focus of this study, the presence of a high degree of metastability in a random network ($10^4$ rewirings) with non-overlapping subsets of oscillators without phase-lagged interaction prompted further investigation. In \hyperref[sec:appendixB]{Appendix B} \hyperref[fig:extra_study]{Fig. 14}, we report an example of a system of oscillators constructed using the configuration model with a uniform degree distribution $k=51$. The random network system is only characterized by the presence of a partition of oscillators, similar to the partition in layer $1$ we studied here, in which only pairs of oscillators that do not belong to the same subset $c$ in the partition have phase-lagged interaction. Interestingly, results confirm the presence of metastable chimera states very similar to those displayed in the Shanahan model \cite{Shanahan2010} despite the absence of heterogeneous couplings and modular structures.
\newline

With the addition of natural frequencies, the system's coupling becomes an important factor in the dynamics. It is well known that oscillatory systems display various degrees of metastability within the range of critical couplings \cite{Villegas2014}. However, we seek to investigate the behaviour of the system away from the critical synchronization point to assess in what range of values of $\delta\omega$ the results we reported in the previous section with $K=50/k$ can hold with the addition of heterogeneous frequencies. In the case of $k=51.2$ ($k=21$), the analysis reported in \hyperref[fig:Results4_part2]{Fig. 9} shows that for values of $\delta\omega<2$ ($\delta\omega<10$) metastable dynamics at all layers remain unchanged. As the heterogeneity of frequencies is increased further, all blocks of oscillators in each layer display incoherent dynamics as each average local KOP $R_i$ reaches the inverse of the square root of the number of oscillator in a layer $1$ block, i.e., $\lim_{\delta\omega\gg\langle\omega\rangle}R_i=\left(\sum^i_k n_{k}\right)^{-1/2}$ (for instance, $R_3$ reaches $1/\sqrt{n_1}=1/\sqrt{16}$). Consequently, the indexes of metastability for high frequency heterogeneities reflect this incoherence. Peaks in metastability are also detected at values of $\delta\omega$ for which the slope of decreasing $R_i$ is maximal (dashed red lines in \hyperref[fig:Results4_part2]{Fig. 9}), in good agreement with previous studies \cite{Villegas2014}.

\section{Discussion}\label{sec:discussion}
One of the hallmarks of complex systems is the presence of hierarchically modular structures: systems composed of interrelated substructures whose interaction across scales is thought to be fundamental for robust and efficient information processing \cite{Simon1991}. The identification of interrelated subsystems and their role in shaping the dynamics and in the emergence of macroscopic collective patterns is an important ongoing research endeavour in complexity science \cite{Jensen2023} and particularly in network neuroscience \cite{Bassett2011}. In this work, we hypothesised that the interplay between nested structures alone can give rise to chimera and metastable states in minimal brain-inspired models of phase-frustrated oscillatory systems. First, we provided evidence for the emergence of symmetry-breaking states in the absence of heterogeneous couplings, delays, and structural heterogeneities. Second, we revealed the relationships between the symmetry-breaking states we observed and the spectral properties typical of hierarchically modular networks.
\newline

Our analysis highlighted two distinct pathways towards achieving the metastable regime in different layers. First, metastable dynamics were detected in layers $2$ and $3$ at a point in which stable and breathing chimeras cease to exist. While it was possible to explain the emergence of stable and breathing chimera states by integrating out the fast modes of the system and reducing the network to a structure similar to that of the two-population model, the precise mechanism that leads to metastable chimera states remains unclear. Due to the small size of the rescaled system, we cannot exclude the possibility that the behaviours we observed are the result of finite size effects. Additionally, a good agreement with the results in \cite{Abrams2008} was only possible for low values of $k$ and the presence of a large second spectral gap. This result suggests that, when the gap between fast and slow modes is small, the contribution from the fast modes should not be discarded. To reach a conclusive answer, an analytical approach is likely necessary. In \cite{Abrams2008}, the two-population model was solved via the Ott-Antonsen dimensionality reduction approach \cite{Ott2008}, by showing the existence of inhomogeneous solutions to the dynamical system of equations in the limit of an infinite number of nodes in each population. The inhomogenous solutions correspond to the stable and breathing chimera states which are very similar to the chimeras observed in our model. However, the two-population model does not predict the emergence of the metastable chimera states we observed for a wide range of the parameter space. Due to the nested nature of the order parameters in our model and the absence of heterogeneous couplings, a similar analytical approach is not immediate. Informed by our observations, we suggest that a synergetic-based study of hierarchical systems of oscillators could be an interesting future research approach \cite{Zheng2024}. As an intuitive example, in Sec. \ref{sec:Results_layer1} we found that chimera states detected in layer $2$ coexist with highly metastable modules in layer $1$. Crucially, these modules display a high average degree of synchronization (\hyperref[fig:results2]{Fig. 5B}) despite the high degree of metastability. This result suggests a synergetic interpretation of the behaviours we observed, in which the spontaneous (de)synchronization patterns of the modules may be interpreted as fast fluctuations of the ``enslaved" fine-grained order parameters of the system \cite{Haken2004}.
\newline

Second, the emergence of spontaneous (de)synchronization patterns in layer $1$ was found to be related to the size of the second spectral gap. When there is enough separation of timescales and modules are structurally well defined (high modularity), each oscillator within a module has time to integrate with other units in the same module the phase-frustrated inter-module interactions. When the separation of timescales is not large enough, however, modules do not have time to integrate those interactions, resulting in the spontaneous desynchronization of oscillators within the module. This behaviour has been observed in Sec. \ref{sec:Results_layer1} and \ref{sec:Results_Laplacian}, in which we systematically increased the average of the network, leading to the loss of modularity in layer $1$, as well as in Sec. \ref{sec:Results_robustness}, in which modularity was lost by randomly rewiring edges in the network. We also considered the more parsimonious hypothesis that spontaneous (de)synchronization is caused by increasing the average degree of the network since this increases the number of phase-frustrated interactions in the system. However, as the average degree of the network is increased beyond a threshold value, layer $1$ modules do not display increasing metastability; instead metastability reaches a plateau for all degrees $k > 30$ (see \hyperref[sec:appendixB]{Appendix B} \hyperref[fig:metastability_layer_1_var_k]{Fig. 13}).
\newline

Our systematic numerical analysis across different levels of observation allowed us to detect and compare metastable dynamics and chimera states in different hierarchical layers. However, it also highlighted the limitations of the variation of the whole-system KOP as a measure of metastability. For example, while we detected metastability at the whole-system level when the structural parameter ranged between $0.3$ and $0.49$, layer $2$ analysis revealed the presence of stable and breathing chimera states instead. At the same time, metastable dynamics in layer $1$ remained entirely undetected in our analysis at the whole-system level. In computational neuroscience, the variation of the whole-system KOP is often used to tune whole-brain dynamical models, and has been hypothesized to peak when the correlation with dFC is maximised \cite{Deco2016}. In the model we studied, and which allows a more systematic analysis of the indexes of metastability due to the model's inherent modular construction, the peaks in metastability were all found to be in the same region of the parameter space, except for the value of metastability of the modules in the unstable population (\hyperref[fig:results2]{Fig. 5C}). Therefore, our findings suggest that a thorough analysis across scales should be performed when possible, and that the index of metastability should be properly defined in terms of the relevant order parameters of the system. This approach may help resolve inconsistencies in the current literature, where maximising the variation of the whole-system order parameter has not necessarily aligned with the best empirical fit \cite{Pope2023}.
\newline

When dealing with empirical structural brain networks, such analysis is often not possible due to the lack of well defined communities and consequently the absence of clearly identified relevant local order parameters of the system. Both in our model and the Shanahan model \cite{Shanahan2010}, where the metastability index was originally formulated, the systems exhibit a clear modular structure, inherent to the system's construction. Consequently, the issue of identifying the relevant local KOPs is absent. Novel approaches based on the harmonic decomposition of structural connectomes \cite{Atasoy2016, Atasoy2017} have been proposed (see also the eigen-microstate approach \cite{Zheng2024}), which we suggest could overcome this problem. Within this framework, dynamical functional connectivities can be characterized by computing the contribution of each eigenmode of the system at each point in time \cite{Luppi2024, Zheng2024}. Inspired by the harmonic brain modes framework and the work of Villegas and colleagues \cite{Villegas2014, Villegas2022, Villegas2023}, we analysed our systems using the spectral information of the graph Laplacian. Our analysis in layer $1$ showed that when the relative separation of timescales between modes is large enough the fast eigenmodes associated with dynamics in the first layer do not contribute to the dynamics of the system, since modules remain completely synchronized at all times. In layer $2$, instead, we found that the slow modes of the system encode the relevant information about the system's behaviours at higher coarse-grainings. In particular, the time-rescaled system reduced to the two-population model for a specific range of parameters. These results highlight the effectiveness of the LRG approach to identify the critical mesoscale structures with a characteristic timescale, and suggest that a similar approach may be used to identify the relevant order parameters of the system at different levels of observation, as well as the spectral profiles that may facilitate hierarchical integration \cite{Wang2021} in oscillatory systems.
\newline

The limitations imposed on the system have allowed us to study the effects of the presence of a mesoscale structure in isolation, however, it is unclear if the behaviours we observed would persist with the addition of structural heterogeneities such as core-periphery structures, and, in particular, the presence of hubs. The rich club has been shown to change the path towards synchronization \cite{GmezGardees2007} when compared to random networks: rather than small clusters synchronizing first, the subset of hubs and the nodes more strongly connected to them synchronize first. The combination of rich club and modular structure has also been investigated in \cite{Arenas2007, ZamoraLpez2016}, highlighting the role hubs play in the formation of isolated synchronized communities and their contribution to the overall functional complexity. It remains unclear how these structural heterogeneities affect systems of phase-lagged or phase-delayed oscillators and, in particular, away from the critical regime. As noticed in \cite{Arenas2007}, the dynamical equation corresponding to a hub node is a topological average of the phases of its neighbours. In the presence of frustration, it is unclear if the frustrations get cancelled out, promoting integration and synchronization, or have the opposite effect. The answer to this will likely depend on the choice between uniform or heterogeneous lags, or, with the addition of an underlying metric space, heterogeneous delays.
\newline

Finally, we believe that the study of metastability in hyperbolic networks \cite{Krioukov2010} of oscillators might help address these limitations. Hyperbolic networks have been shown to display many of the universal properties of real-world networks \cite{Krioukov2010}, including clustering, hubs, rich club, and modularity \cite{Kovcs2021, Balogh2023}. Crucially, hyperbolic networks come equipped with an underlying metric space, allowing a geometric renormalization procedure \cite{GarcaPrez2018} which has been shown to detect the possible organizing principles of the brain \cite{Zheng2020} across scales.

\printbibliography

\newpage
\section{Appendix A: Hierarchical Network Model Derivation}\label{sec:AppendixA}
We modeled the probability of connection in the upper layers as $p_2(H) = (\frac12 + \frac{H}{2})\gamma$ and $p_3(H) = (\frac12 - \frac{H}{2})\gamma$,  where the factor $\gamma$ in both expressions is used to control the average degree of the system. Then, for $H = 0$ we have $p_2(0) = p_3(0) = \gamma/2$ and for $H = 1$ we obtain $p_2(1) = \gamma$ and $p_3(1) = 0$. Given these constraints and using $n_3 = 2$ as in our model, we may write the average degree of each node as:
\begin{eqnarray}
    \langle{k}\rangle &=& (n_1-1)p_1(H) + (n_2 - 1)n_1p_2(H) + (n_3 - 1)n_1n_2p_3(H) \\
    &=& (n_1-1)p_1(H) + (n_2 - 1)n_1p_2(H) + n_1n_2p_3(H) \\
    &=& (n_1-1)p_1(H) + (n_2 - 1)n_1\left[\left(\frac{1+H}{2}\right)\gamma\right] + n_1n_2\left[\left(\frac{1-H}{2}\right)\gamma\right] \\
    &=& (n_1-1)p_1(H) + n_1n_2\gamma - \frac{H + 1}{2}n_1\gamma,
\end{eqnarray}
where in the first step we used $n_3 = 2$ which we used in this study. By requiring the derivative of $\langle k\rangle$ with respect to $H$ to be zero we obtain
\begin{eqnarray}
    \frac{\mathrm{d}\langle{k}\rangle}{\mathrm{d}H} &=& (n_1-1)\frac{\mathrm{d}p_1}{\mathrm{d}H} - \frac{n_1}{2}\gamma = 0 \\
    \frac{\mathrm{d}p_1}{\mathrm{d}H} &=& \frac{n_1}{2(n_1-1)}\gamma \\
    \Rightarrow p_1 &=& \frac{n_1}{2(n_1-1)}\gamma H + C.
\end{eqnarray}
Note that $C$ is arbitrary. However, since $p_1\in[0,1]$, $H\in[0,1]$, and we required $p_1$ to be as close to unity as possible, we may define $C$ to be equal to
\begin{equation}
    C = 1 - \frac{n_1}{2(n_1-1)}\gamma,
\end{equation}
without loss of generality. Finally, we rewrite $p_1$ as
\begin{equation}
    p_1(H) = 1 - \left(\frac{n_1 \gamma}{n_1 - 1}\right)\left(\frac{1 - H}{2}\right).
\end{equation}
\newpage

\section{Appendix B}\label{sec:appendixB}
\subsection{Sensitivity Analysis}
\begin{figure}[h!]
    \centering
    \includegraphics[width=1.0\textwidth]{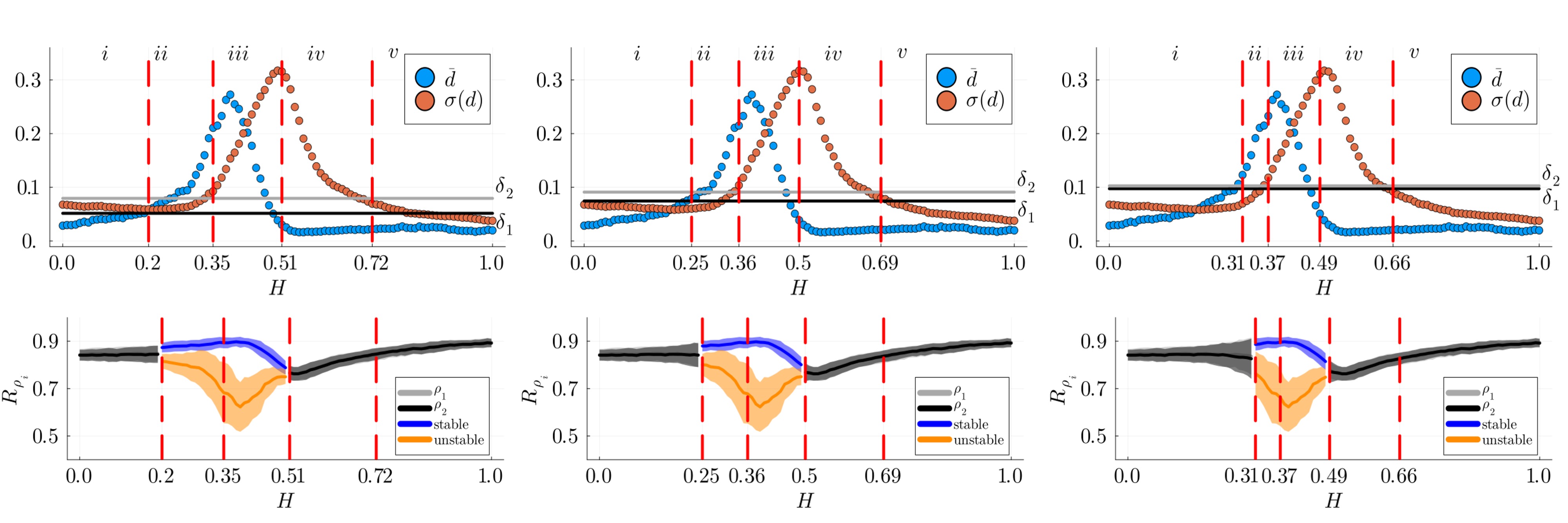}
    \caption{Top row: Average and standard deviation of the difference between the populations' local KOPs, $\bar{d}$ and $\sigma(d)$ respectively, as a function of $H$ for fixed $k=51.2$. The threshold values $\delta_1,\,\delta_2$ were chosen to be $s$ standard deviations higher than the baseline values obtained for $H=0.0$. Left, middle and right panels show the regions identified when using $s=1$, $s=2$ and $s=3$ respectively. Whilst the boundaries change, the results in \hyperref[sec:results1]{Sec. 3.2} do not qualitatively change (bottom row).}
    \label{sec:fig_sensitivity}
\end{figure}

\begin{figure}[h!]
    \centering
    \includegraphics[width=1.0\textwidth]{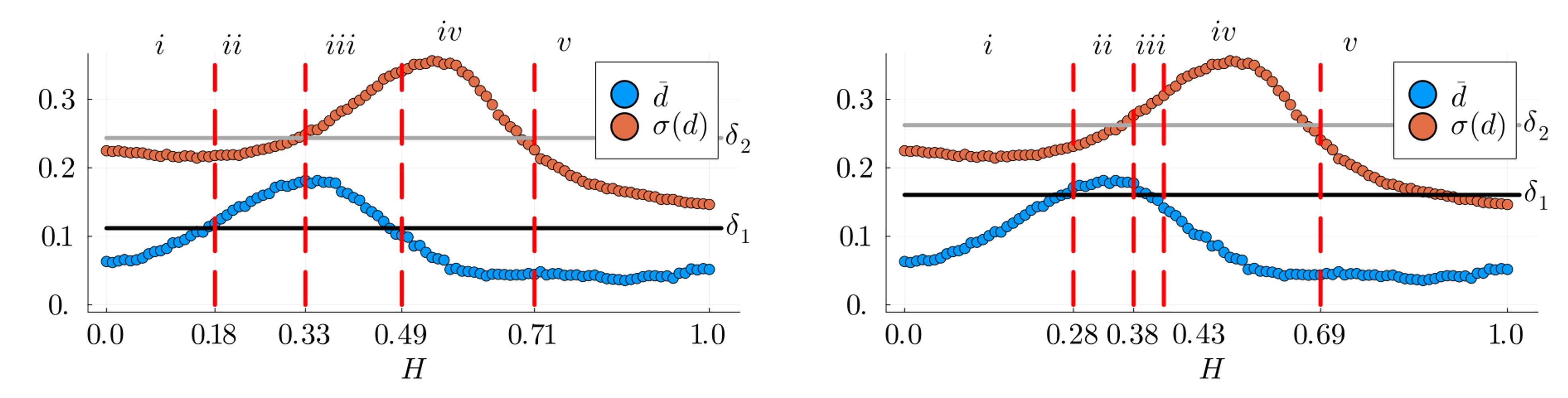}
    \caption{Average and standard deviation of the difference between the populations' local KOPs, $\bar{d}$ and $\sigma(d)$ respectively, as a function of $H$ for fixed $k=21$. The threshold values $\delta_1,\,\delta_2$ were chosen to be $s$ standard deviations higher than the baseline values obtained for $H=0.0$. Left and right columns show the regions identified for $s=1$ and $s=2$, respectively. For higher values of $s$ stable and breathing chimeras are not detected.}
    \label{fig:identify_chimeras_k_21}
\end{figure}

The regions of the parameter space in which chimera states are detected is found by computing the average and standard deviation of the difference between the populations' local KOPs across a simulation. We report the values we found as a function of $H$ averaged over $100$ realizations for fixed $k=51.2$ in \hyperref[sec:fig_sensitivity]{Fig. 10}, where we show how the regions limits vary depending on the choice of the arbitrary threshold while leaving our results qualitatively invariant.
\newpage

\subsection{Alternating Chimera States}
\begin{figure}[h!]
    \centering
    \includegraphics[width=1\textwidth]{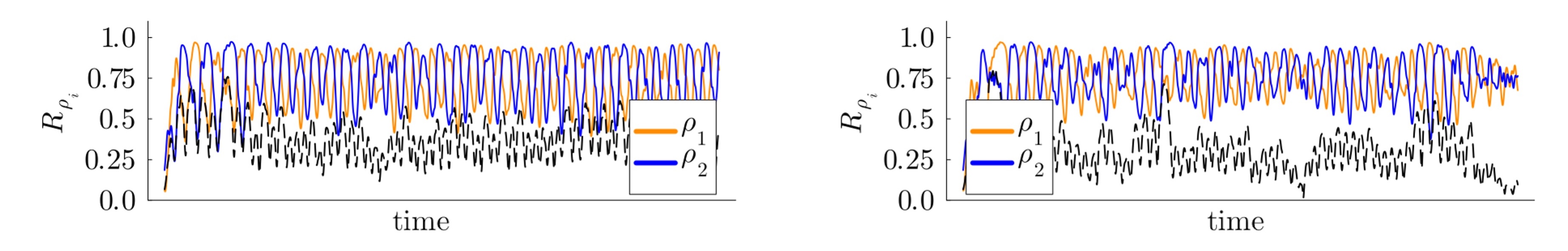}
    \caption{Alternating chimera can be found for values of $H>0.5$, for instance $H=0.55$ (left) and $H=0.58$ (right).}
    \label{fig:alternating}
\end{figure}

\subsection{Metastability as a function of the average degree}
\begin{figure}[h!]
    \centering
    \includegraphics[width=1.0\textwidth]{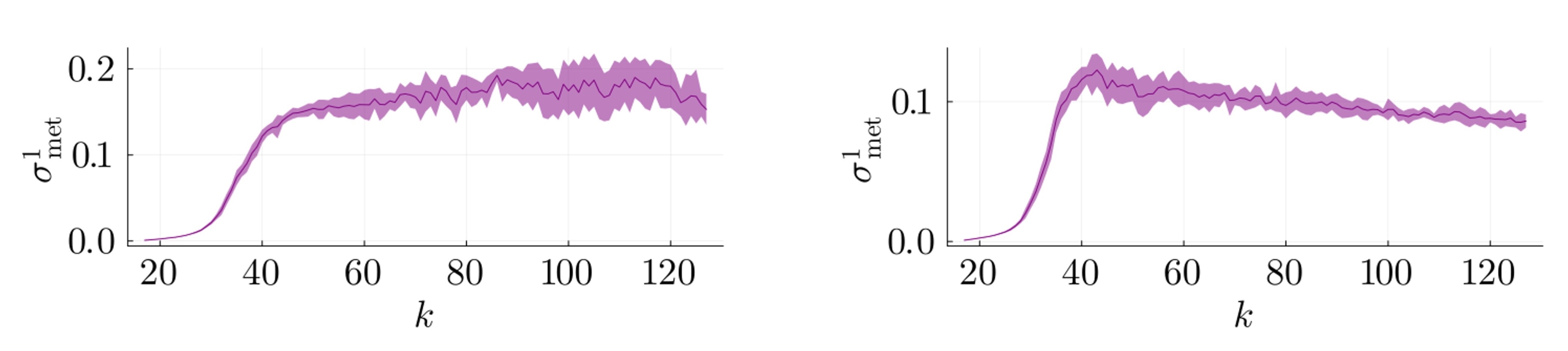}
    \caption{Metastability in layer $1$, $\sigma^1_{\mathrm{met}}$, as a function of $k$ for fixed $H=0.5$ (left) and $H=0.3$ (right).}
    \label{fig:metastability_layer_1_var_k}
\end{figure}

\subsection{Configuration model with phase-lags}
\begin{figure}[h!]
    \centering
    \includegraphics[width=1.0\textwidth]{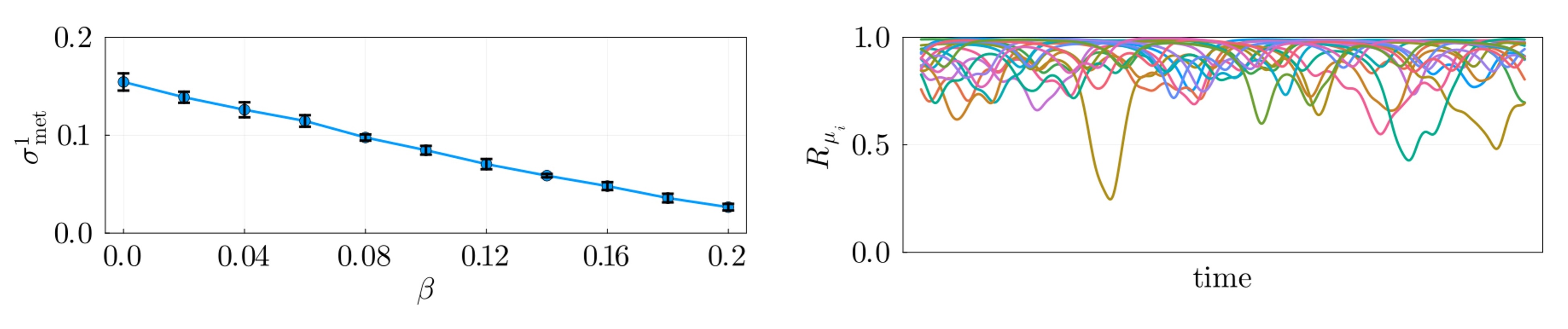}
    \caption{Left: Metastability as a function of the lag parameter in random networks of oscillators constructed using the configuration model with average degree $k=51$. Only pairs of oscillators which are associated with a different subset $c$ as defined by the partition vector $P_1$ share a phase-lagged interaction with phase-lag $\alpha=\pi/2 - \beta$. Right: example of modules' local KOP dynamics in time. Simulations performed using Euler method with step size $10^{-3}$ for $25000$ steps. Metastability values and error bars obtained from $10$ random initialization.}
    \label{fig:extra_study}
\end{figure}

\end{document}